# 3D- ($H$-$\theta$-$\varphi$) magnetic phase diagram of antiferromagnetic metal $GdB_6$ with electron and lattice instability

A. N. Azarevich[1], A. V. Bogach[1], T. F. Garipova[1], V. V. Voronov[1], M. Rajnak[2], S. Gabani[2], K. Flachbart[2], N. B. Bolotina[1,3], O. N. Khrykina[1,3], V. M. Gridchina[3,1], A. Yu. Tsvetkov[4], S. Yu. Gavrilkin[4], N. E. Sluchanko[1]

*[1]Prokhorov General Physics Institute, Russian Academy of Sciences, Vavilov str. 38, Moscow 119991, Russia*

*[2]Institute of Experimental Physics of the Slovak Academy of Sciences, Watsonova 47, SK-04001 Košice, Slovakia*
*[3]National Research Center "Kurchatov Institute", Academician Kurchatov sq., 1, Moscow 123182, Russia*
*[4]Lebedev Physical Institute, Russian Academy of Sciences, Leninsky Av. 59, Moscow 119991, Russia*

* *e-mail: nes@lt.gpi.ru*

**Abstract.** The origin of charge transport and magnetization anisotropy was studied in $GdB_6$, an antiferromagnetic (AF) metal (Néel temperature $T_N \approx 15.5$ K) with cubic lattice and Gd S-type magnetic ions. Both small static Jahn-Teller distortions and nanoscale electronic instabilities (dynamic charge stripes) were found in precise low temperature X-ray diffraction measurements. The detailed magnetic field ($H$) vs temperature ($T$) phase diagrams were constructed with two main magnetic phases AF(I) and AF(II). Using the angular $\varphi$-dependences of magnetoresistance and magnetization, *impeller-type* patterns of the $H$-$\varphi$ magnetic phase diagrams in the (110) and (111) planes were found at helium temperatures, which included the AF phases I and II separated from each other by radial and circular boundaries. The results argue in favor of the important role of the spin density wave 5d- component in the magnetic structure of AF(II) state. Charge fluctuations in stripes are proposed to be responsible for the suppression of the Ruderman-Kittel-Kasuya-Yoshida (RKKY) indirect exchange between the first and second neighboring $Gd^{3+}$ ions located along the <100> and <110> directions. These dynamic charge stripes and vibrationally coupled Gd-Gd pairs produce unusual anisotropy of charge scattering and the *impeller-type* diagrams in $GdB_6$ with S-type magnetic ion.

## I. Introduction

In the family of rare earth (RE) hexaborides $RB_6$ with a simple cubic CsCl-type crystal structure, there are several exotic antiferromagnetic (AF) metals that demonstrate unusual magnetic ordering with the unique wave vector $\boldsymbol{q}_M$= (¼, ¼, ½) in presence of charge density modulation. Indeed, very similar AF ground state was reported for the archetypal heavy fermion antiferromagnet $CeB_6$ [1-2], as well as for $PrB_6$ in the low-temperature commensurate AF phase [3-5], for $GdB_6$ [6-10], $TbB_6$ [10-12], $DyB_6$ [13-14] and $HoB_6$ [15]. In most of these RE borides the complex magnetic structures are attributed usually to quadrupole (multipole) inter-site interactions, but this is not the case, at least for $GdB_6$. In this hexaboride trivalent Gd ion has a half-filled 4*f*-shell without orbital degrees of freedom ($4f^7$, $L$= 0), and the order at $\boldsymbol{q}_M$ should have an origin, which does not depend so much on the particular configuration of 4f-electrons [16-17]. Kuramoto and Kubo [16-17] have argued that the wave vector $\boldsymbol{q}_M$ must reflect peculiarities of the Fermi surface constructed from 5d-2p electron states of RE-ion and boron, and these features are common to all AF rare-earth hexaborides. Thus, the physical origin of the complex AF magnetic structures found in the ground state of $RB_6$ remains controversial, and $GdB_6$ appears to be a suitable candidate to shed more light on the mechanism of AF ordering at $\boldsymbol{q}_M$= (¼, ¼, ½) in presence of distortions in the cubic lattice.

$GdB_6$ undergoes two successive phase transitions with simultaneous structural distortions and AF ordering [18, 6-10]. Below Néel temperature $T_N \approx 15.5$ K, a magnetic structure characterized by the propagation vector $\boldsymbol{q}_M$= (¼, ¼, ½) [6-10] and a crystal structure modulation by $\boldsymbol{q}_{L1}$= (½, 0, 0) simultaneously arise as a result of the first-order phase transition. Amara et al. proposed the magnetoelastic coupling mechanism by taking into account very small (0.004-0.016 Å) displacements of Gd ions [9], which led to a dependence of the RKKY-exchange integral on the distance between the relevant magnetic moments. This modulated crystal structure is slightly suppressed with cooling at $T^* \sim 9$ K, and other charge modulations associated with $\boldsymbol{q}_{L2}$= (½, ½, 0) are superimposed with $\boldsymbol{q}_{L1}$ while maintaining the same $\boldsymbol{q}_M$ magnetic structure, which corresponds to both the atomic displacement and the magnetic ordering [18, 8-9]. It is worth noting that, according to a resonant X-ray scattering study of $GdB_6$, the satellite reflections created by $\boldsymbol{q}_M$= (¼, ¼, ½) also exhibit anomalies in the amplitude at $T^* \approx 9$ K, but no other magnetic scattering wave vector was found [7]. The $\boldsymbol{q}_{L2}$ charge modulation is expected to be more stable than the $\boldsymbol{q}_{L1}$ modulation owing to the strong electron–phonon coupling characterized by the phonon kink at $\boldsymbol{q}_k$ = (0.38, 0.38, 0) that is close to $\boldsymbol{q}_{L2}$, being a signature of the cubic lattice instability.

The precursor of the heavy RE-ion large amplitude motion relevant to the structural instabilities was revealed in $GdB_6$ [19, 20], $TbB_6$ [21] and $DyB_6$ [22]. It has been found in [19-22], that the RE-ion motion exhibits phonon dispersion relations located in the energy range 5-10 meV. In particular, a longitudinal phonon propagating along the cubic [100] axis bends down, approaching the Brillouin zone boundary at $\boldsymbol{q}_{L1}$= (½, 0, 0), which is the propagation vector of the distorted structure below $T_N$. The measured energies at $\boldsymbol{q}_{L1}$ decrease by 10–30% with decreasing temperature from 300 K to the antiferromagnetic ordering at $T_N$, and the kink anomalies observed at $\boldsymbol{q}_k$ = (0.38, 0.38, 0) in both longitudinal and transverse modes. The observed phenomena indicate the interplay between the conduction electrons and the motion of guest RE ions filling oversized lattice cavities of boron atoms. The softening of the $\boldsymbol{q}_{L1}$ mode with decreasing temperature increases in the order of atomic numbers of Gd, Tb, and Dy. This phenomenon is explained by the enlargement of the free space for the RE-ion motion inside the $B_{24}$ polyhedron due to the lanthanide contraction.

As a common mechanism responsible for the large-amplitude vibrations (rattling modes) of RE-ions loosely bound to the rigid covalent boron framework in $RB_6$, the authors of [23-27] suggested the development of cooperative Jahn-Teller (JT) instability of the boron sub-lattice, which was established by precise X-ray diffraction measurements of $GdB_6$ [23], $LaB_6$ [24-25] and $CeB_6$ [25-28]. In particular, both JT small static distortions and strong dynamic charge stripes along some of <100> directions in $GdB_6$ lattice were observed at temperatures 85-300 K [23], and it was accompanied with formation of vibrationally coupled Gd-Gd pairs in the stripe directions. A room temperature measurements of wide range dynamic conductivity allowed detecting strong collective modes (overdamped oscillators) in $Gd_xLa_{1-x}B_6$ [29-30], $CeB_6$ [31], $YB_6$ and $YbB_6$ [32]. It has been shown in [29-32], that non-equilibrium (hot) electrons participating in the formation of the collective JT modes dominate in the charge transport of $RB_6$, and smaller fraction of Drude-type electrons changes in the range 25-45% in the studied hexaborides. From the strong-coupling perspective, the dynamic charge stripes are a real-space pattern of micro-phase separation [33], and there is a strong concurrence between these charge fluctuations and the Ruderman–Kittel–Kasuya–Yosida (RKKY) oscillations of spin density of conduction electrons. Therefore, it seems natural to expect (i) the suppression of indirect RKKY exchange between nearest neighbor Gd-ions, and (ii) the breakdown of cubic magnetic symmetry in $GdB_6$, both induced by the charge fluctuations in stripes. Note, that for a number of magnetic RE-dodecaborides $RB_{12}$ ($R$- Dy, Ho, Er, Tm and Yb) similar effects have been established recently in [34-38], leading to complicated AF magnetic phase diagrams with numerous magnetic phases and phase transitions.

Thus, the aim of the study is twofold. Firstly, the results of the detailed measurements of resistivity, magnetization and heat capacity were collected and analyzed to reconstruct magnetic $H$-$T$, the angular $H$-$\varphi$ and three-dimensional (3D) $H$-$\theta$-$\varphi$ phase diagrams of AF state in $GdB_6$ demonstrating a drastic symmetry lowering in this compound with a simple cubic lattice and $S$-type magnetic ions. Secondly, we show the low temperature charge density distribution deduced by the maximum entropy method (MEM) from the precise XRD study of $GdB_6$ at $T$=30 K and 85 K. Two configurations of dynamic charge stripes are considered: (i) formed on the hybridized 5d-2p states of Gd and B and directed along the [110] axes; (ii) located predominantly on the 2p orbitals of boron and directed along the [100] axes. Arguments are presented in favor of the role of these fluctuating charges in magnetic symmetry breaking, low-temperature charge modulation, and the formation of an anisotropic antiferromagnetic ground state in $GdB_6$.

## II. Experimental details

Detailed studies of magnetization ($M$) and resistivity ($\rho$) in combination with heat capacity ($C$) and Seebeck coefficient ($S$) measurements were performed on high-quality single-domain single crystals of $GdB_6$ that were grown by induction zone melting in an inert gas atmosphere [39]. The resistivity measurements were carried out with the help of original setup that allowed the samples to be rotated stepwise around the current axes $\boldsymbol{I}||[001]$, $\boldsymbol{I}||[110]$ and $\boldsymbol{I}||[111]$ in external magnetic field up to 80 kOe. The transverse configuration ($\boldsymbol{I}\perp\boldsymbol{H}$) is applied, when $\boldsymbol{H}$ lies respectively in the (001), (110), and (111) planes containing principal directions in the cubic lattice. Location of the boundaries on the field-temperature ($H$-$T$) and field-angle ($H$-$\varphi$) magnetic phase diagrams ($\angle\varphi = \boldsymbol{H}$^$\mathbf{n}$, where $\mathbf{n}$ is the normal vector to the lateral surface of the sample; see sketch of the sample rotation in Fig. 8a below) was confirmed by the measurements of magnetization, heat capacity and Seebeck coefficient, carried out in the magnetic fields up to 50 kOe and 90 kOe, using the Quantum Design MPMS-5 and PPMS-9 installations, correspondingly. The MPMS-5 was used also to measure the angular dependences of magnetization for $\boldsymbol{H}$ changes in the planes (001), (110) and (111). The sample mount setup provides high precision (~1%) absolute magnetization values, which consider the demagnetization factor, the finite sample size and the radial displacement inside the SQUID-magnetometer pickup coils. Measurements of the magnetization field and temperature dependences in magnetic fields up to 180 kOe were performed using vibrating sample magnetometer with cryo-free 180 kOe magnet (Cryogenic Limited, UK). To estimate small JT static distortions of the cubic lattice and to visualize the dynamic charge stripes in $GdB_6$, a precise X-ray diffraction (XRD) study was carried out on single-crystal samples smaller than 0.15 mm in size. The XRD data set (with Ag$K_\alpha$ radiation, $\lambda$ = 0.56087 Å) was obtained at low temperatures $T$ = 30 K and 85 K using a Rigaku XtaLAB Synergy-DW (Oxford Diffraction) diffractometer with a HyPix-Arc 150° photon detector. To cool the sample, the N-Helix 800 Series cryosystem with an open stream of cold helium directed at the sample was applied. The temperature stability was within ~0.1 K.

## III. Results and discussion

**III.1. *H*-*T* magnetic diagrams of $GdB_6$ for principal *H* directions**. The AF(I)-P (P-paramagnetic state) phase transition at $T_N \approx 15.5$ K in $GdB_6$ is clearly visible on the temperature dependences in small field magnetization at $H_0$=100 Oe, zero field resistivity and Seebeck coefficient, and specific heat in the magnetic field up to 90 kOe (Fig. 1). The second magnetic transition at $T^* \sim 9$ K (I-II transition in Fig. 1) is observed only on the magnetization and resistivity curves. Within experimental accuracy, we did not fixed any significant changes of both the heat capacity and thermoelectric power near $T^*$.

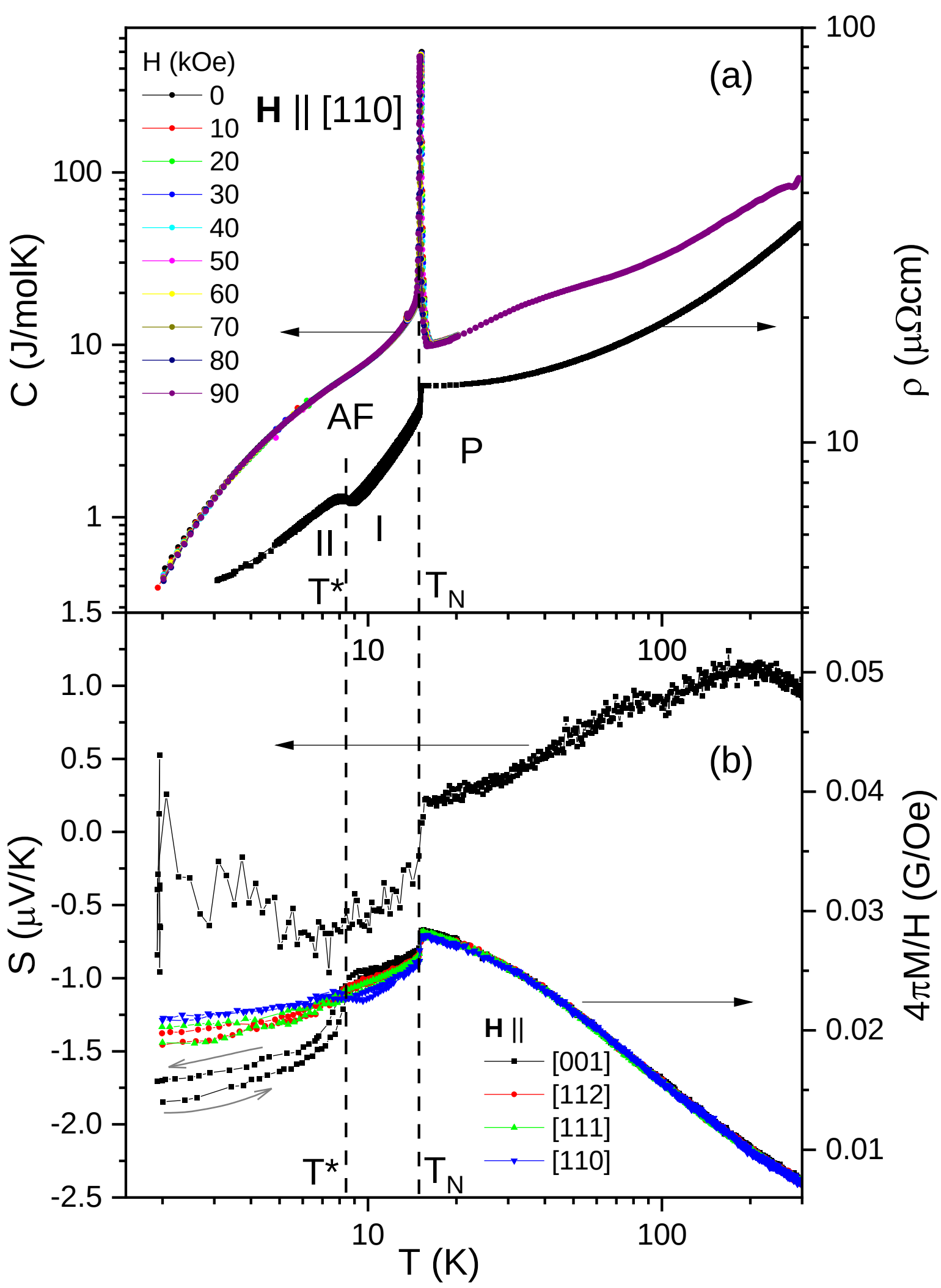

**Fig. 1.** Temperature dependences of (a) the specific heat $C(T, H_0)$ in various magnetic fields $H_0 \leq$ 90 kOe, and zero field resistivity ρ($T$), and (b) the Seebeck coefficient $S(T)$ and magnetic susceptibility $4\pi M/H(T, H_0 = 100$ Oe) for ***H***||[001], [112], [111] and [110] in $GdB_6$. I, II and P denote two AF and paramagnetic phases, $T_N$ and $T^*$ are the temperatures of magnetic phase transitions.

It is worth noting, that the low-field magnetization is about isotropic in the paramagnetic phase (see curves for ***H***||[001], ***H***||[110], ***H***||[111] and ***H***||[112] in Fig. 1b), a moderate magnetic anisotropy appears below $T_N$ in phase I, and it increases below $T^* \approx 8.5$ K in the ground state (phase II, see also Figs. S1 and S2 in the Supplementary Materials [40]). The temperature dependences of magnetic susceptibility, resistivity and Seebeck coefficient demonstrate a step-like decrease at $T_N$ under cooling, which is accompanied with a huge lambda-type anomaly of the specific heat (Fig. 1), and all these singularities are attributed to a weak first-order AF(I)-P phase transition with simultaneous structural distortions and antiferromagnetic ordering [18, 6-10]. A zero field resistivity $\rho(T)$ curves show strong cooling-heating hysteresis in the temperature range $T^*$- $T_N$ (AF phase I in Fig. 1), which disappears only when magnetic field along [110] and [111] directions increases above 20 kOe, but a complicated hysteresis of resistivity is maintained in the case of $\boldsymbol{H}$||[001] (see Fig. 2). Besides, the second magnetic transition I-II is also accompanied with a strong field hysteresis of resistivity in wide vicinity of $T^*$ (Fig. 2).

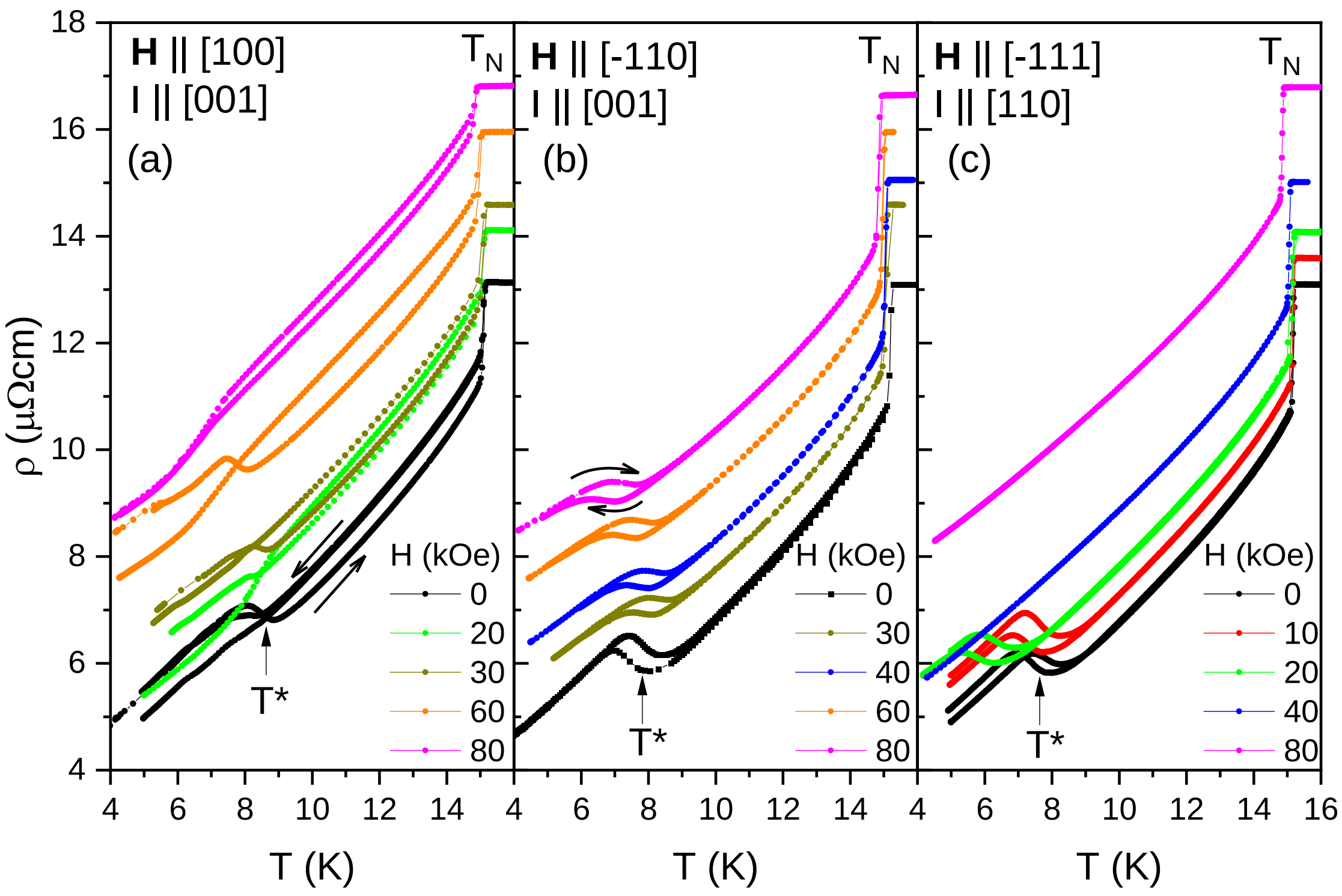

**Fig. 2.** Temperature dependences of resistivity in the external magnetic field $H \leq 80$ kOe directed along (a) [100], (b) [-110] and (c) [-111]. For clarity, the curves on the panels are shifted upward by 0.5 μΩ·cm for every 10 kOe.

A hysteresis near $T_N$ and $T^*$ is observed also on the zero-field cooled – field cooled temperature dependences of susceptibility $4\pi M/H_0(T, H_0)$ for a principal directions $\boldsymbol{H}$||[100], $\boldsymbol{H}$||[110] and $\boldsymbol{H}$||[111] (see, for example, Figs. 3a-3d, and also Figs. S3-S5 in [40]). Additionally, Fig. 3a shows the $4\pi M/H_0(T, H_0)$ curves for $\boldsymbol{H}$||[100] up to 180 kOe, indicating magnetic anomalies near $T_N$ and $T^*$. Fig. 3d demonstrates field dependences $4\pi M/H(H, T_0)$ for $\boldsymbol{H}$||[111] with two hysteresis area limited by $H_m$ and the critical field $H^*(T)$ separating the AF phases I and II. Field dependences of resistivity recorded at various temperatures in the range $T_0 \leq 16$ K are shown in Fig. 4. It is discerned that the hysteresis of resistivity $\rho(H_0, T_0)$ (Figs. 2 and 4) and susceptibility $\chi(H_0, T_0)$ (Fig. 3, see also Figs. S3-S5 in [40]) appears synchronously, and additional anisotropic phase boundary $H_m(T)$ was detected for $\boldsymbol{H}$||[111] below 20 kOe, dividing each of phases I and II into two regions. A similar lower field phase boundary at about 5 kOe is observed also for direction $\boldsymbol{H}$||[110], separating low field hysteresis areas $I_h$ and $II_h$ from high field regions I and II (see Fig. S5 in [40]). The same magnetic hysteresis is found also for $\boldsymbol{H}$||[100] in the interval $H \leq$ 3.5 kOe ($II_h$ phase in Fig. 5 below) and it may be attributed to the remagnetization of antiferromagnetic domains. The anomalies detected on the field and temperature curves of magnetic susceptibility (Fig. 3 and Figs. S1-S5 in [40]) and resistivity (Figs. 2 and 4), allowed us to establish the phase boundaries between $II_h$, II, $I_h$, I and P states, and reconstruct a magnetic $H$-$T$ phase diagram of $GdB_6$ along principal directions in the cubic lattice (Fig. 5). Note, that the $H$-$T$ diagram (Fig. 5) is more detailed than those detected in [41-43], including additional low-field branches found in the AF I and II phases.

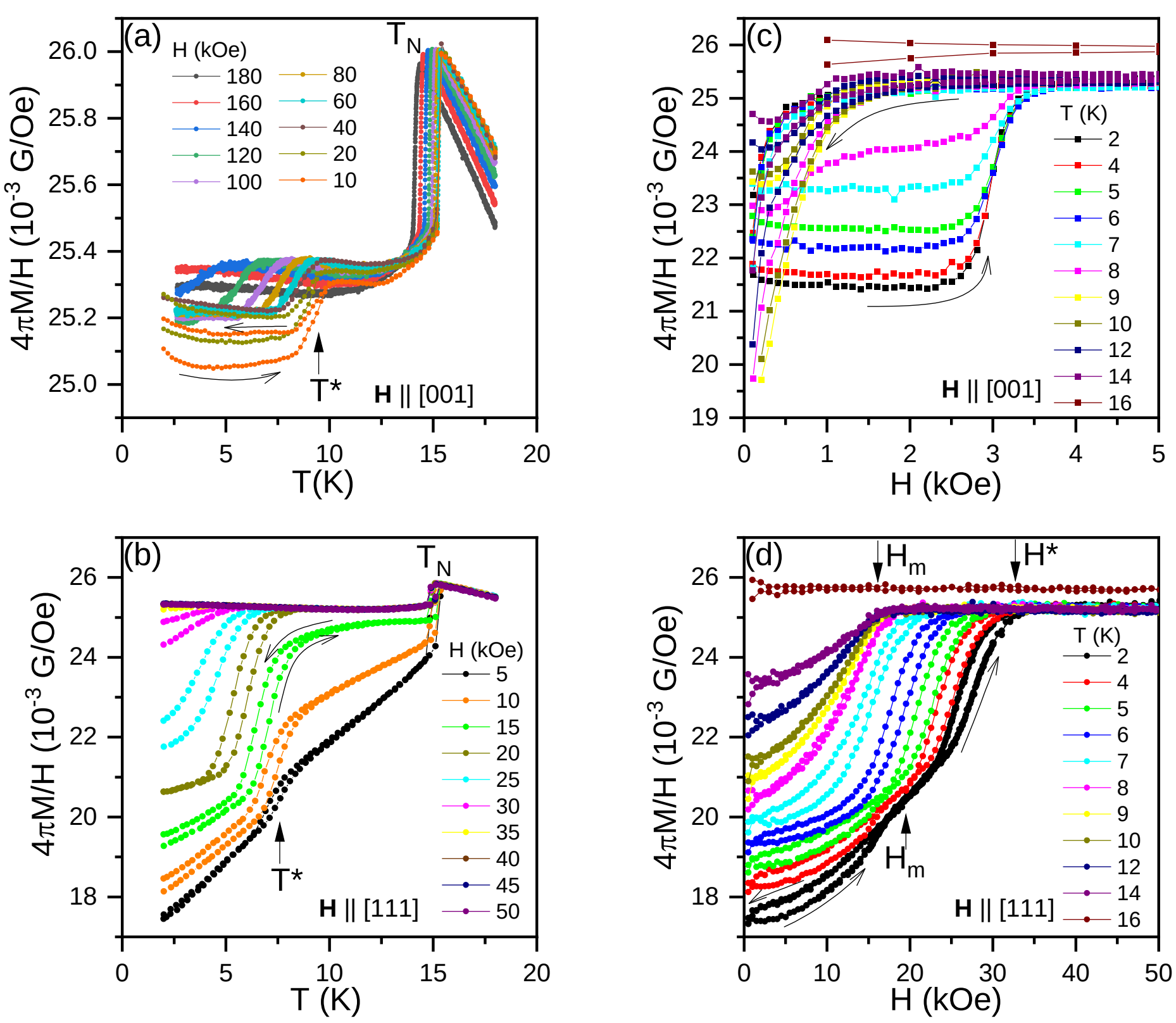

**Fig. 3.** Temperature dependences of susceptibility $4\pi M/H_0(T, H_0)$ of $GdB_6$ (a) in the range $T \leq 18$ K in magnetic field up to 180 kOe for the principal direction $\boldsymbol{H}||[001]$, and (b) in magnetic field up to 50 kOe for $\boldsymbol{H}||[111]$. Panels (c) - (d) show field dependences of the susceptibility at temperatures fixed in the range 2-16 K for (c) $\boldsymbol{H}||[001]$ in the interval $H \leq 5$ kOe, where the hysteresis is observed, and for (d) $\boldsymbol{H}||[111]$ up to 50 kOe. Vertical arrows mark the magnetic phase transitions I-II and I-P.

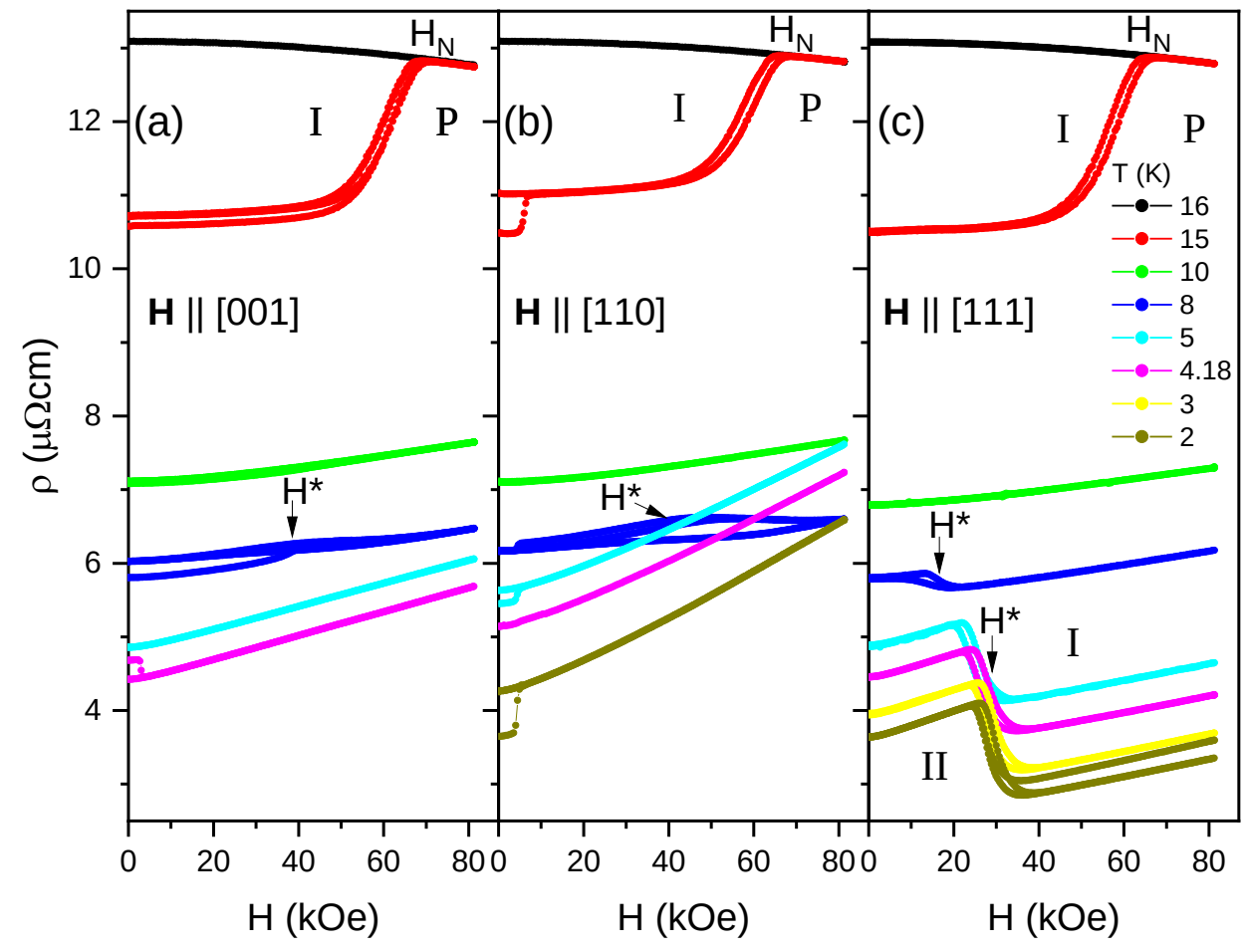

**Fig. 4.** Magnetic field dependences of resistivity $\rho(H, T_0)$ of $GdB_6$ at fixed temperatures $T_0 \leq 16$ K for principal field directions $\boldsymbol{H}||[001]$ (a), $\boldsymbol{H}||[110]$ (b) and $\boldsymbol{H}||[111]$ (c). Arrows mark the magnetic phase transitions in the AF state, Roman numerals denote the magnetic phases (see Fig. 5). $H_N$ and $H^*$ denote the critical fields of the AF phase transitions at $T_N$ (I-P) and $T^*$ (I-II).

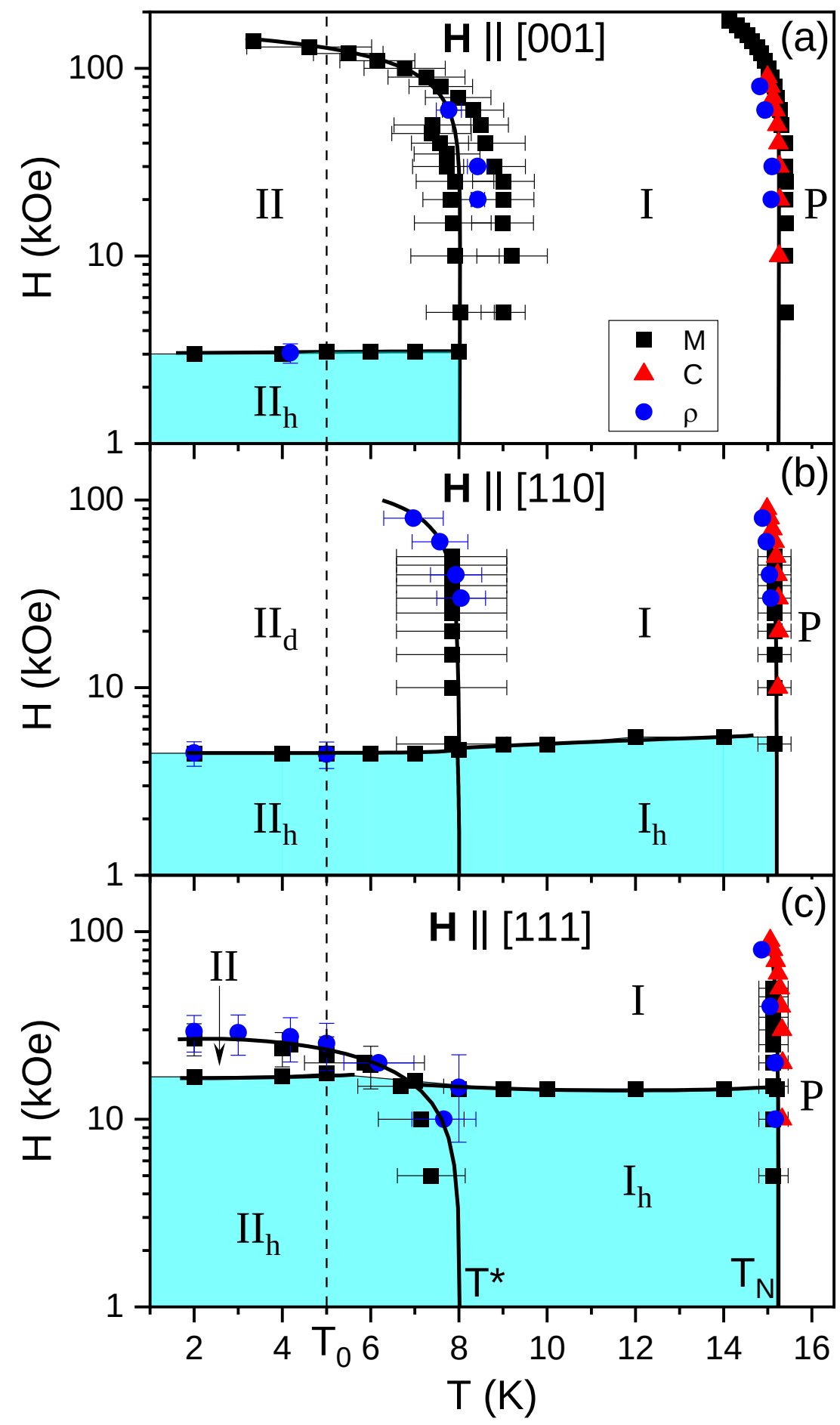

**Fig. 5.** Magnetic *H*-*T* phase diagrams of $GdB_6$ along three principal directions of the external magnetic field (a) ***H***||[001], (b) ***H***||[110] and (c) ***H***||[111]. Roman numerals denote different magnetically ordered phases, P- paramagnetic state. Vertical dashed line marks $T_0 = 5$ K, which corresponds to the results of angular measurements (Figs. 6-7) and to the $H$-$\varphi$ diagram in the (110) and (111) planes (Figs. 8-9). Low field hysteresis areas (phases $I_h$ and $II_h$) are shaded.

Three important issues of Figs. 1-5 should be point out. (1) Both the charge transport and magnetic anisotropy are small enough in the range 9-15 K in the AF I phase (see also Figs. S6 and S7), and they appear and increase below $T^*$, when field-induced II–I phase transition occurs for field direction ***H***||[111] (see Figs. 2-5). (2) A sharp climbing (hump) on the resistivity curves $\rho(T)$ under cooling just below $T^*$ is due to the involvement of conduction electrons in collective states of charge and spin density waves (CDW and SDW), arising in the complex AF ground state in this magnetic metal (see, for example, similar anomalies caused by itinerant antiferromagnetic phase in chromium [44]). As a result, in $GdB_6$ the main difference in the magnetic structure of the AF I and II phases with a single propagation vector $\boldsymbol{q}_M$= (¼, ¼, ½) [7] should be attributed to emergence of additional 5*d*- (SDW) component in the AF ground state (phase II). (3) A strong magnetic hysteresis is observed in the phase II for the magnetic field direction ***H***||[111], which may be attributed to significant disordering of Gd magnetic ions embedded in the oversized cavities in the rigid boron cage ($B_{24}$ complexes, see Fig. 14a below). Note, that similar SDW 5d- component of magnetic structure has been discussed recently in the series of cage-cluster RE dodecaborides with loosely bound state of RE ions,- $Ho_xLu_{1-x}B_{12}$, $ErB_{12}$, $TmB_{12}$ and $DyB_{12}$ in [45, 34-38]. The hump on the resistivity curves of $GdB_6$ at $T^*$ is suppressed completely in the external magnetic field above 30 kOe directed along [111] axis (field-induced II-I phase transition, Figs. 2c and 5), but the $\rho(T, H_0)$ anomaly is still observed in the strong magnetic field ***H***||[001] and ***H***||[110] (Fig. 2a, 2b). The behavior may be attributed to suppression of SDW component of magnetic structure for ***H***||[111].Similar field-induced anisotropy of resistivity and magnetic susceptibility was established recently in $DyB_{12}$, $Ho_xLu_{1-x}B_{12}$, $ErB_{12}$ and $TmB_{12}$ [45, 34-38], resulting in the Maltese Cross and the butterfly-type angular magnetic phase diagrams for ***H*** located in the (110) plane. Following the approach developed in [45, 34-38] for studies of magnetic anisotropy in the $RB_{12}$ antiferromagnetic metals, the results of angular resolved resistivity and magnetization measurements of $GdB_6$ are presented and discussed below.

### III. 2. Angular *H*-*φ* magnetic diagrams in the (001), (110) and (111) planes at $T_0$ = 5 K.

Firstly, to study the magnetic anisotropy in the AF phases of $GdB_6$, the angular dependences of the magnetic susceptibility $4\pi M/H(\varphi, H_0, T_0)$ have been investigated at different temperatures $T_0 \le 20$ K. In these experiments the magnetic field vector $\boldsymbol{H}_0$ was located in a {110} family plane. The values of $H_0$ were fixed at1 kOe, 5 kOe and 18 kOe (see panels (a), (b) and (c) of Fig. 6, correspondingly; the color shows the amplitude of the magnetic signal). The plane from the {110} family was chosen because the vector **H** lying in this plane can coincide with each of the three principal directions in a rotating cubic crystal. As it was mentioned above, the most significant magnetic anisotropy in the Néel phases appears below $T^*$~ 9K (phase II in Fig. 6), the magnetic signal decreases noticeably in vicinity of [111] direction for $H_0$< 20 kOe (hysteresis areas in Fig. 5), which may be attributed to a strong critical spin fluctuations near the radial AF phase boundaries. A complicated magnetic anisotropy appears at $H_0$= 18 kOe below $T^*$ (phase II in Fig. 6c) where the $II_h$-II transition is observed in wide vicinity of ***H***||[111] direction (see Fig. 5c).

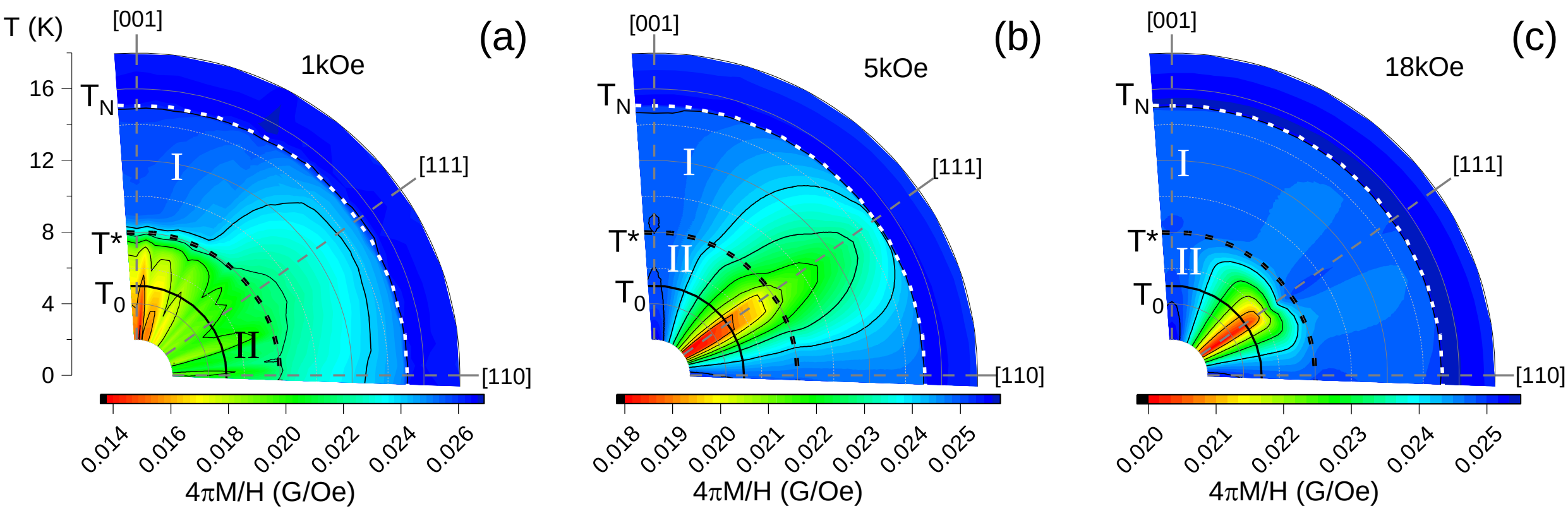

**Fig. 6.** (Color online). Temperature dependences of magnetic susceptibility in the range $T \le$ 18 K for magnetic field intensity $H_0$ = 1 kOe (a), 5 kOe (b) and 18 kOe (c) with rotation of ***H*** in the (1-10) plane. Radial dotted lines show the principal directions when vector ***H*** is aligned with the principal axes in the cubic lattice. Circular dotted lines mark $T_N$, $T^*$ and $T_0$ = 5 K. Roman numerals I and II denote different magnetically ordered phases. The amplitude of $4\pi M/H(\varphi, H_0, T_0)$ is shown by color.

It is seen in Fig. 5 that the location of the AF-P phase boundary at $T_N$ ~ 15 K is similar for various ***H*** orientation, whereas the critical field $H^*(T)$ of AF I-II transition depends dramatically from ***H*** direction. Indeed, at $T_0$ = 5 K the critical field $H^*$ changes from ~25 kOe for ***H***||[111] to ~ 125 kOe are found in the [001] field direction (Fig. 5). To clarify the nature of the strong low-temperature magnetic anisotropy observed in $GdB_6$ crystals, where S-type magnetic Gd ions are arranged in a simple cubic lattice, angular dependences of the magnetic susceptibility $4\pi M/H(\varphi, H_0, T_0)$ and the resistivity $\rho(\varphi, H_0, T_0)$ were studied at a fixed temperature $T_0$ = 5 K ($T_0$ marked by a vertical dotted line in Fig. 5 and dotted circles in Fig. 6). In these series of experiments, the direction and intensity of ***H*** was varied in (001), (110) and (111) planes. For example, Figs. 7 and 8 show the sets of the angular resolved magnetic susceptibility and resistivity dependences, respectively, recorded at various fixed external magnetic fields.

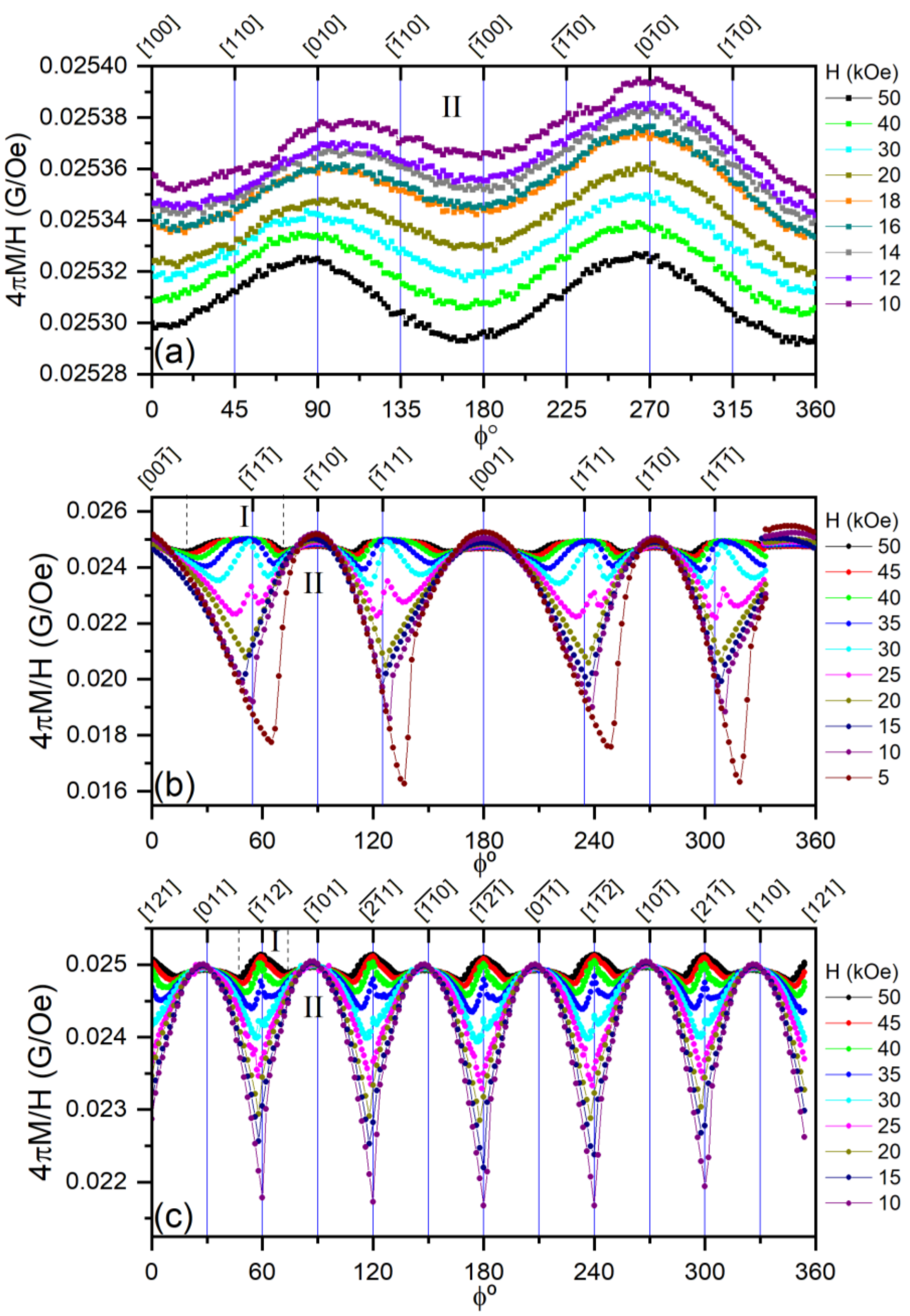

**Fig. 7.** (Color online). Angular dependences of magnetic susceptibility for the various magnetic fields in the planes ***H***||(001) (a), ***H***||(110) (b) and ***H***||(111) (c) at $T_0$ = 5 K. Vertical lines show positions when the magnetic field ***H*** in these planes is aligned with the principal axes in the *cubic* lattice (marked above the top axes). Roman numerals denote the magnetic phases (see phase diagram in Fig. 5).

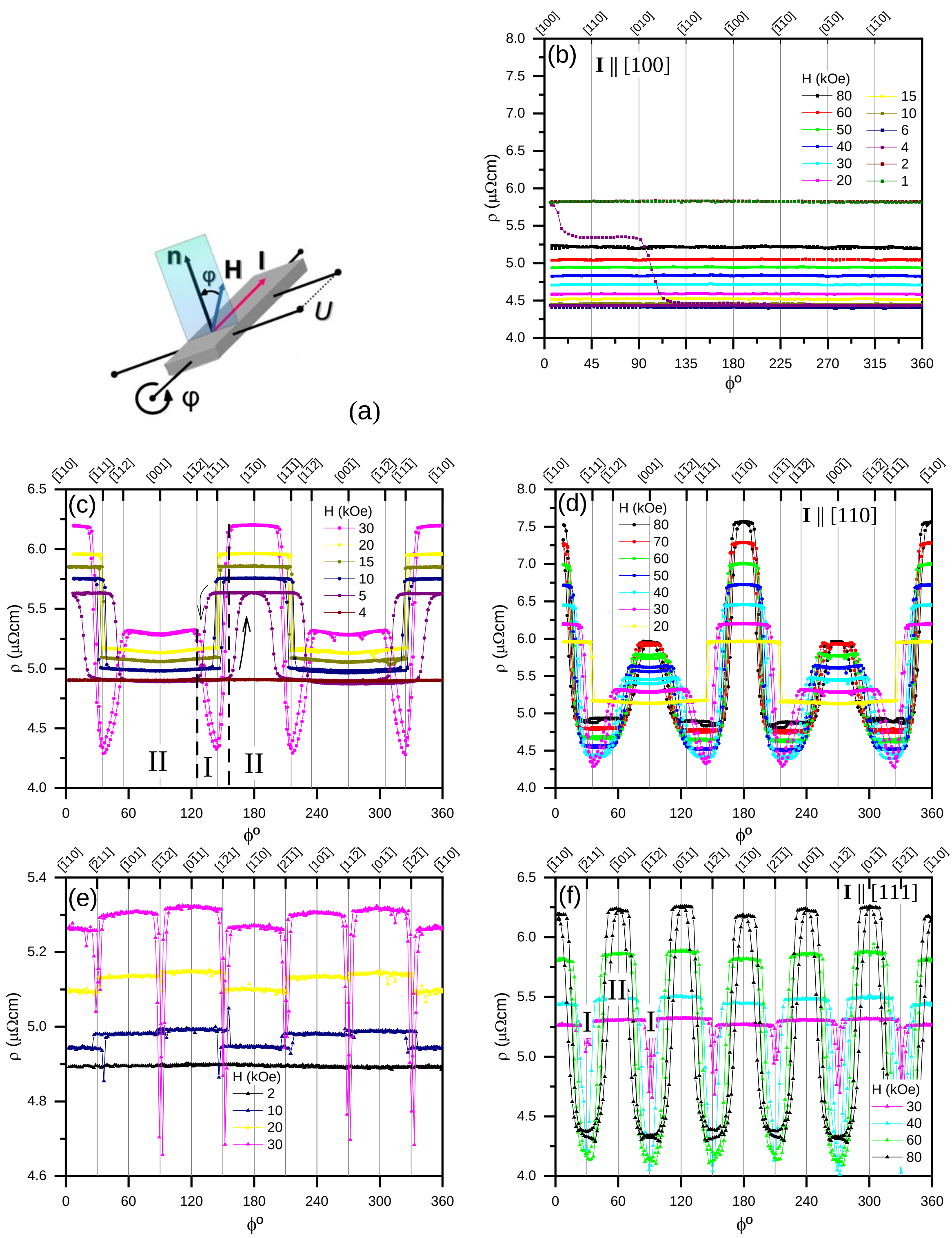

**Fig. 8.** (Color online). (a) Sketch of resistivity experiment with the step-by-step sample rotation. Angular dependences of resistivity for various intensity of transverse (***H***⊥***I***) magnetic field ***H***||( 001), panel (b), ***H***||(110), panels (c)-(d) and ***H***||(111), panels (e)-(f), at $T_0$ = 5 K. Vertical lines show positions when the magnetic field ***H*** in the planes ( 001) (b), (110) (c)-(d) and |(111) (e)-(f) is aligned with the principal axes in the cubic lattice (marked above the top axes). Roman numerals denote the magnetic phases (see Fig. 5).

Additionally, we investigated magnetic field dependences of susceptibility $4\pi M/H(H, \varphi_0, T_0 = 5$ K) and resistivity $\rho(H, \varphi_0, T_0 = 5$ K) for different ***H*** directions, when the vector ***H*** was fixed at various angles $\varphi_0$ in the planes (110), (110) and (111) (see, for example, Fig. 9, angle $\varphi_0$ is measured approximately from the principal direction [-110]; sketch of the sample rotation is shown in Fig. 8a). It is discerned in Figs. 6-9, that (*i*) the direction ***H***||[111] differs from others in the interval $H$ < 25 kOe, exhibiting critical behavior of the measured parameters. (*ii*) Magnetic and charge transport anisotropy appears in the magnetic field $H > H^*(T_0$=5 K) ~25 kOe, increasing strongly above 40 kOe. (*iii*) Sharp and simultaneous changes on the angular dependences of $4\pi M/H(\varphi_0, H_0)$ and $\rho(\varphi_0, H_0)$ allow detecting the location of phase transitions at $T_0$ = 5 K both near the <111> axes and in a wide vicinity of <100> directions (radial and circular phase boundaries in the (110) and (111) planes). (*iv*) Step-like anomalies with a strong hysteresis are observed on the resistivity field dependences in the interval $H > H^*(T_0$ = 5 K) ~ 25 kOe (Figs. 9b and 9c), and it may be attributed to the intersection of circular phase boundaries in the planes (110) and (111). On the contrary, only negligible (~0.1%) itinerant magnetic anisotropy was found in the (100) plane both in the measurements of susceptibility (Fig. 7a) and resistivity (Figs. 8b and 9a).

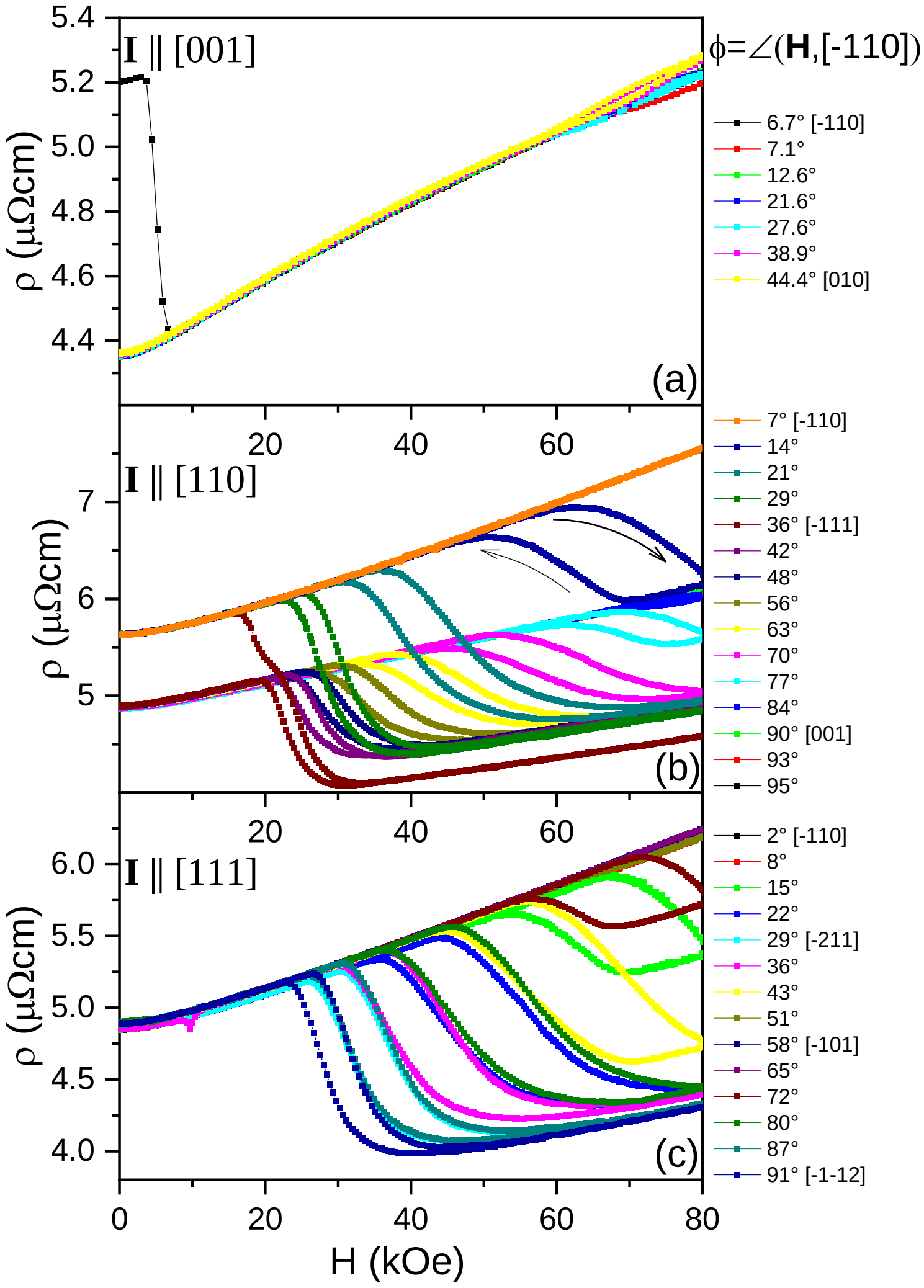

**Fig. 9.** Sweep-up and sweep-down magnetic field dependences of resistivity for the various transverse (***H*** ⊥ ***I***) magnetic fields $H \leq 80$ kOe at $T_0$ = 5 K. The step-by-step sample rotation was performed around the direct current axis (a) ***I***||[100], (b) ***I***||[110] and (c) ***I***||[111] (see sketch in Fig. 8a).

An overview of the results of $\rho(H, \varphi)$ and $4\pi M/H(H, \varphi)$ measurements at $T_0$ = 5 K presented in Figs. 7-9 is displayed in the polar coordinates in Figs. 10-11, where the values of resistivity and magnetic susceptibility are shown using a color scale. The projections of these $\rho(H, \varphi)$ (Figs.

8-9) and $4\pi M/H(H, \varphi)$ (Fig. 7) data sets onto the (110) and (110) planes (see Figs. 10 and 11) allow refining the $H$-$\varphi$ AF phase diagrams at $T_0$ = 5 K. Roman numerals in Figs. 10-11 show the magnetic I and II phases in the AF state (see also Fig. 5). Note, that the “impeller patterns” of the magnetic and charge transport anisotropy in the (110) plane detected here for $GdB_6$ for the first time (Figs. 10a and 11a), looks similar to the “Maltese cross” type picture obtained in the AF phase in $HoB_{12}$ [34] and $TmB_{12}$ [35-36]. Let us note also that, although the rotation of the magnetic field in the (100) plane does not lead to any noticeable changes in resistivity (Fig. 8b, 9a) and magnetization (Fig. 7a), the situation changes significantly when $\boldsymbol{H}$ rotates in the (110) plane.

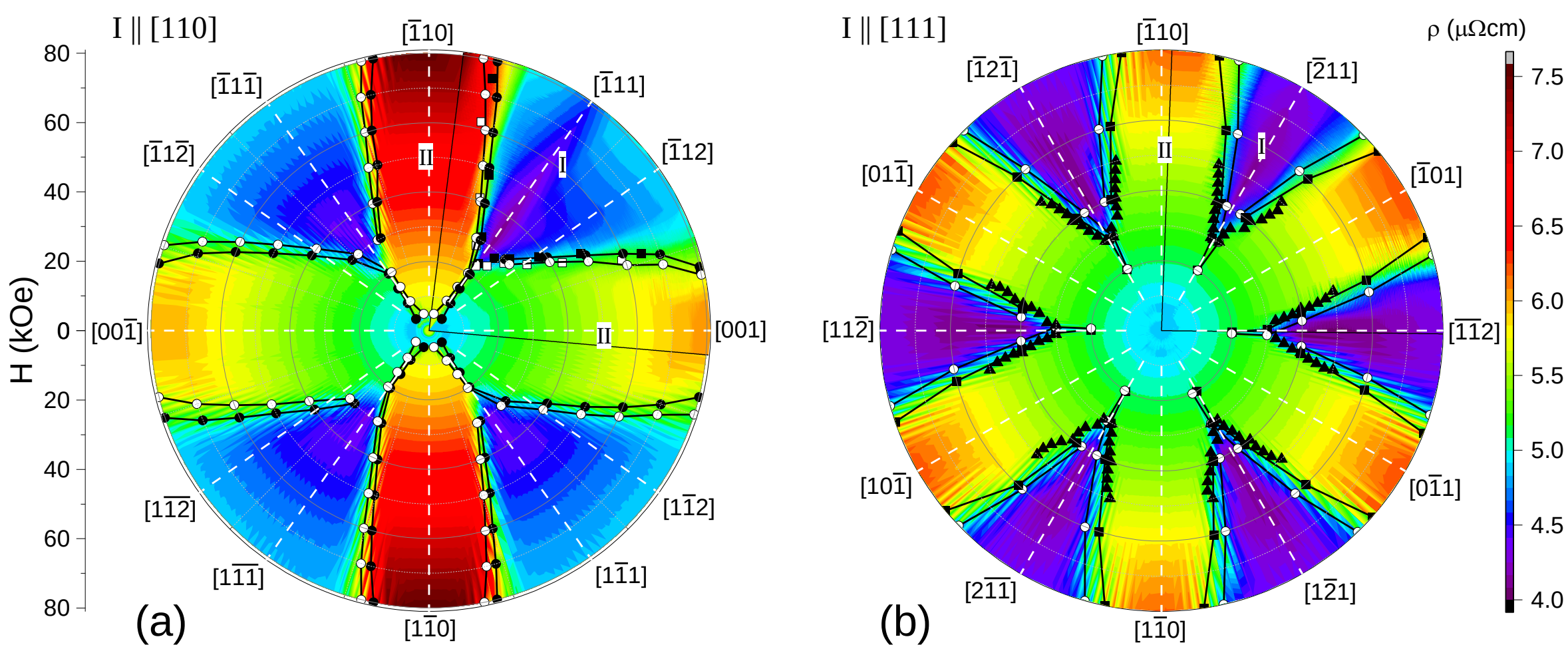

**Fig. 10.** Resistivity of $GdB_6$ in polar ($H$, $\varphi$) coordinates in projection onto the (a) (110) and (b) (111) planes recorded at $T_0$ = 5 K for $H \leq 80$ kOe. Color shows the magnitude of resistivity. Phase boundaries are shown as dots and lines, Roman numerals present different magnetic phases in the AF state, which are identical to those shown in Fig. 5.

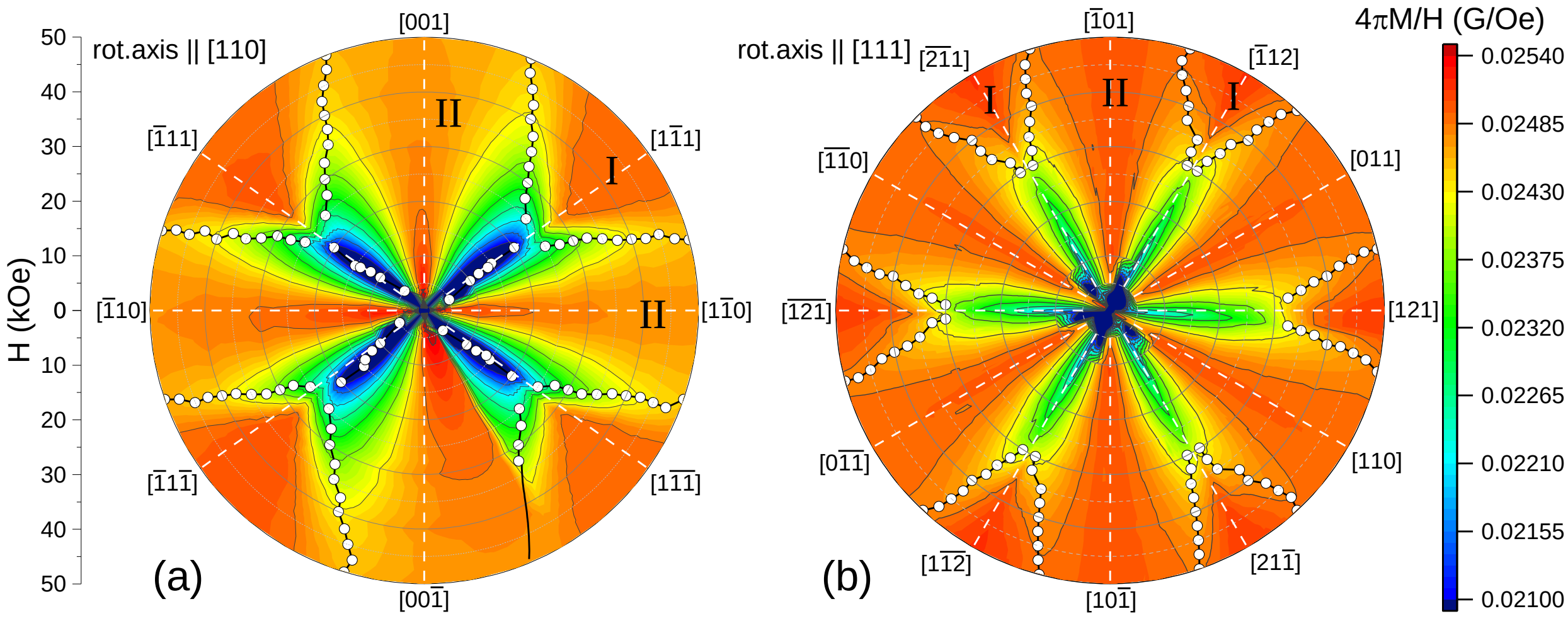

**Fig. 11.** Magnetic susceptibility of $GdB_6$ in polar ($H$, $\varphi$) coordinates in projection onto the (a) (110) and (b) (1-11) planes recorded at $T_0$ = 5 K for $H \leq 50$ kOe. Color shows the magnitude of $4\pi M/H$ parameter. Phase boundaries are shown as dots and lines, Roman numerals present different magnetic phases in the AF state, which are identical to those shown in Fig. 5.

Indeed, rotation of the vector $\boldsymbol{H}$ in the (110) plane, passing through regions near the <111> routes ($\Delta\varphi_{111}$ sectors) with strong hysteresis, leads to a significant increase in the resistivity in the $\Delta\varphi_{110}$ sectors (vicinity of <110> directions, see Figs. 8c-8d). The effect may be attributed to disordering of Gd ions located in the large size cavities of the boron sub-lattice. Thus, one can suppose that the AF phase II is the same for sectors $\Delta\varphi_{001}$ and $\Delta\varphi_{110}$ in $GdB_6$ (see Figs. 10a, 11a) and the first order transition through the <111> phase boundaries leads to the disorder-induced resistivity increase for $\boldsymbol{H}$ directions located in the $\Delta\varphi_{110}$ sector.

**III. 3. Angular 3D $\boldsymbol{H}$-$\boldsymbol{\varphi}$-$\boldsymbol{\theta}$ magnetic diagram at $\boldsymbol{T_0}$ = 5 K.** The borders on the planar $H$-$\varphi$ diagrams (Fig. 10) provide important information related to the arrangement of two different sections of the *spherical* $H$-$\varphi$-$\theta$ (3D) *magnetic phase diagram*. Fig. 12 shows the schematic view of the combined $H$-$\varphi$-$\theta$ diagram as the projection on the spherical surfaces with $H_0$ = 40 kOe (panel (a)) and 80 kOe (b) of these basic high-field phases I and II (the areas colored by deep-blue and red, correspondingly). Solid lines on the spherical surfaces in Fig. 12 indicate the angular $\varphi = (\mathbf{n}, \boldsymbol{H})$ trajectories of the vector $\boldsymbol{H}$, associated with three experiments of the resistivity measurements. These correspond to the sample rotation: (1) between $\boldsymbol{H}$||[ 100] and $\boldsymbol{H}$||[ 010] in the (001) plane (see Fig. 8b); (2) from $\boldsymbol{H}$||[001] to $\boldsymbol{H}$||[110] passing through $\boldsymbol{H}$||<111> directions in the (1-10) plane (see Fig. 8d); and (3) from $\boldsymbol{H}$||[011] to $\boldsymbol{H}$||[1-10] passing through $\boldsymbol{H}$||<112> directions in the (11-1) plane (see Fig. 8f).

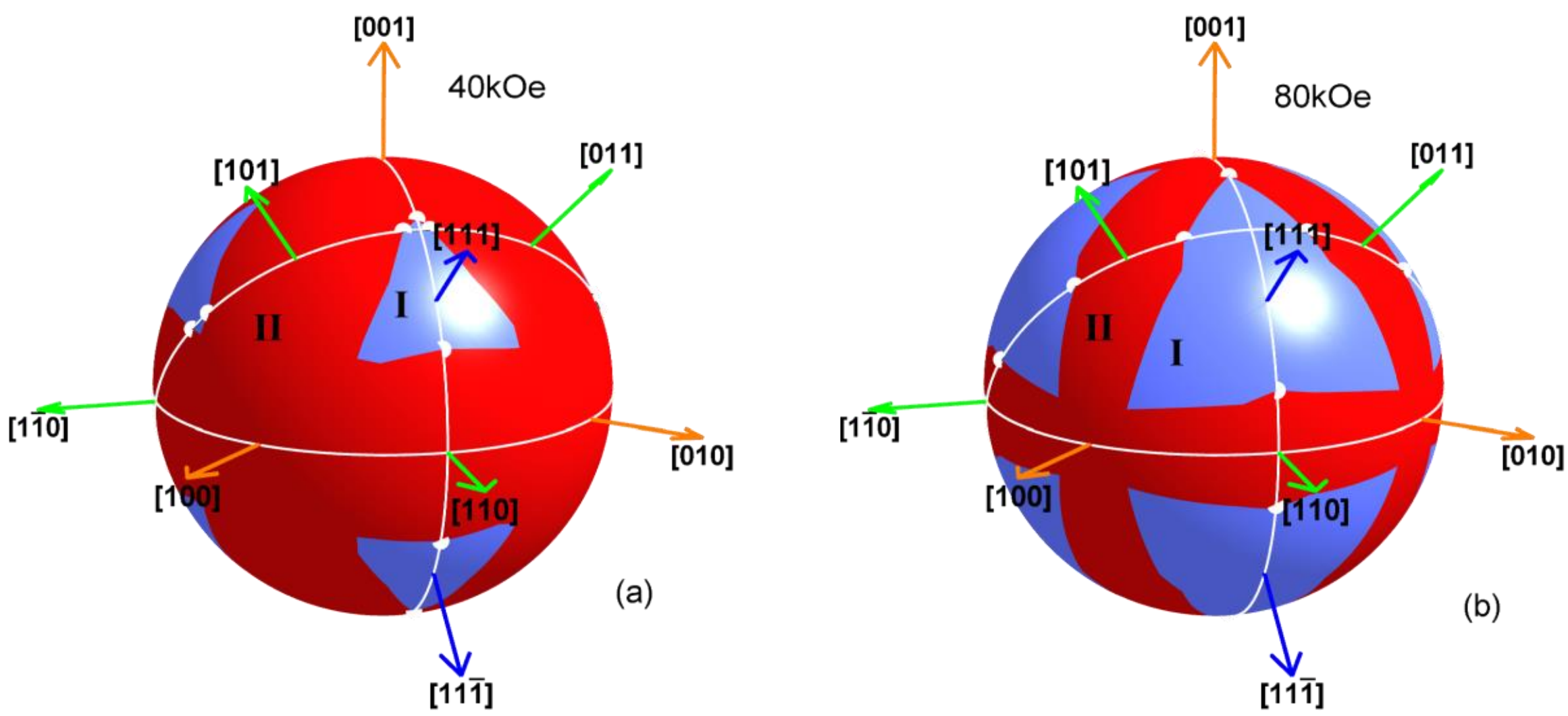

**Fig. 12.** (a) Schematic view of the projection on the spherical surfaces $H_0$ = 40 kOe (a) and 80 kOe (b) of two high-field phases I and II in the $H$-$\varphi$-$\theta$ phase diagram constructed at $T_0$ = 5 K. The white solid lines indicate the trajectories of the vector $\boldsymbol{H}$ in the crystal if the latter rotates around the crystallographic axes [001], [1-10], [11-1], which are also the direct current axes. Phase boundaries are shown as dots and lines on the red-blue borders, Roman numerals present magnetic phases I and II in the AF state, which are identical to those shown in Fig. 5.

**III. 4. Analysis of magnetoresistance components.** Anomalies at $T_0$= 5 K on the angular dependences of magnetic susceptibility $4\pi M/H(\varphi_0, H_0)$ and resistivity $\rho(\varphi_0, H_0)$ appear in the same directions and at the same intensity of the external magnetic field (see, Figs. 7-11), which demonstrates synchronous changes in the two characteristics. These singularities should be attributed to orientation-dependent magnetic phase transitions. Field-induced increase of resistivity $\rho(\varphi_0, H)$ in the AF phases I and II is about linear both in phase II and above the II-I phase transition (Fig. 9 and Fig. S8 in [40]). Above 30 kOe the resistivity and therefore magnetoresistance (MR) $\Delta\rho/\rho(\varphi_0, H) = \rho(\varphi_0, H)/\rho(\varphi_0, H{=}0)-1$ becomes strongly anisotropic distinguishing three main angular segments $\Delta\varphi_{001}$, $\Delta\varphi_{110}$ and $\Delta\varphi_{111}$ located around three principal crystallographic directions [001], [110] and [111] (Fig. 13) and separated one from another by abrupt radial and circular borders (Figs. 10, 12). Slight deviations from the linear MR behavior are observed both in the small-field hysteresis zone $H < 5$ kOe (Figs. 8c, 8e) and above 20 kOe in the narrow vicinity of [001] direction (Fig. 9a). These deviations may be attributed to the appearance in the sector $\Delta\varphi_{001}$ of a small negative quadratic MR component, which is reliably recorded on the field dependence of the derivative $d(\Delta\rho/\rho)/dH(H)$ (see Fig. S9 in [40]).

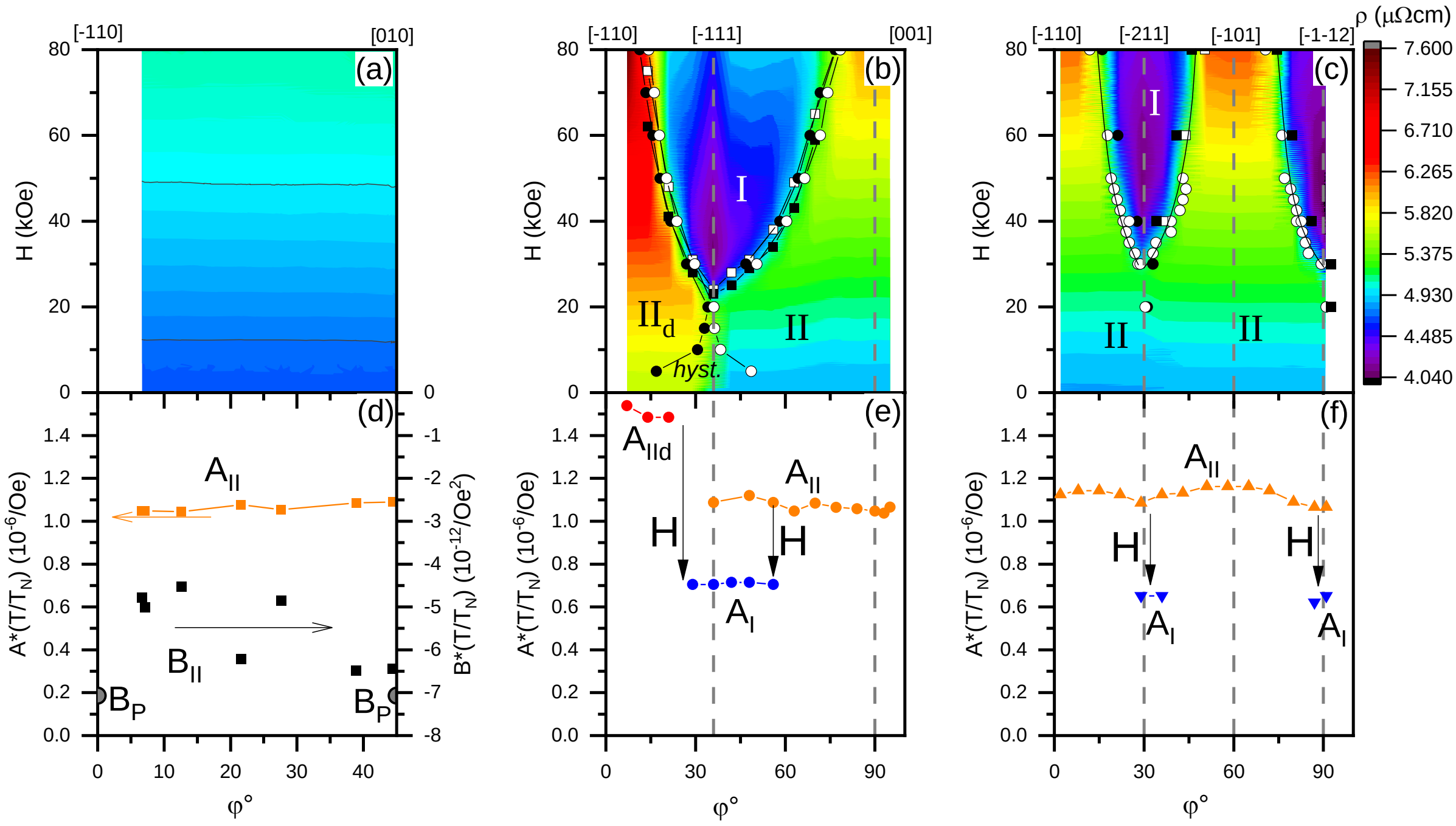

**Fig. 13.** (a) Field dependences of resistivity in the ($H$, φ) coordinates in projection onto the planes ( 001) (a), (110) (b) and (111) (c) (color shows the magnitude of resistivity, phase boundaries are indicated by dots and lines). Panels (d), (e) and (f) show the linear positive $A_{I,II}$ coefficients of magnetoresistance in the planes ( 001), (110) and (111), correspondingly (see text), detected in the phases I and II at $T_0$ = 5 K. Panel (d) shows also quadratic negative coefficient $B$ found at $T_0$ = 5 K in the AF(II) ($B_{II}$) and in the paramagnetic state at $T_0$ = 16 K ($B_P$). Roman numerals denote magnetic phases I, II and II$_d$ in the AF state, which are identical to those shown in Fig. 5.

Accordingly, following the approach to the MR data analysis developed in [38], we used approximations:

$$\frac{\Delta\rho}{\rho} = \begin{cases} A_{II}H + \frac{B_{II}}{2}H^2, & \text{when } H \in \Delta H_{II} \\ A_I(H - H^*), & \text{when } H \in \Delta H_I \end{cases} \qquad (1),$$

where $H^*(T)$ is the critical field separating the AF phases I and II, and $A_{I,II}$ and $B_{II}$ are the linear positive and quadratic negative coefficients in the magnetoresistance terms. Here we take into account that in sector $\Delta\varphi_{111}$ of AF phase I the changes of MR at $T_0$ = 5 K should be measured from the critical field $H^*$. Note, that positive $A_{I,II}$ coefficients are associated usually with the charge carriers scattering on spin-density waves, whose antinodes are considered as a 5$d$ component of the complex antiferromagnetic structure, composed of magnetic moments arranged on the 4f and 5d electron states [34, 45]. As shown in [46-47], SDW component of magnetic structure becomes more intensive in moderate external magnetic field, providing a linear increase of positive MR. The scattering of charge carriers on the SDW in the AF state decreases gradually with the sample heating, the lowest $A_I(T)$ values are detected just below $T_N$, and negative quadratic MR ($B_P$) dominates usually above $T_N$ (Fig. 4). Besides, over the entire temperature range, the inequality $A_{II} > A_I$ is valid for $\mathbf{H}\|[111]$ (Fig. 4c), i.e., in this field direction the amplitude of charge carriers scattering on SDW decreases sharply at $H^*(T)$ indicating SDW weakening in the AF phase I.

The analysis of the MR contributions (see Eq.(1)) developed in various magnetic phases of $GdB_6$ at $T_0 \approx 5$ K allows us to deduce and compare the angular changes of $A_{I,II}$ and $B_{II}$ coefficients (see Fig. 13) with the location of phase boundaries on the angular ($H$-$\varphi$) magnetic diagrams in the planes (001) (panels (a)-(d)), (110) ((b)-(e)) and (111) ((c)-(f)). It is seen in Fig. 13, that within the AF phases I, II and disordered phase $II_{dis}$ (Fig. 5) the $A_{I,II}$ and $B_{II}$ values change very slightly being the characteristics of two scattering mechanisms discussed above, and behavior of the MR coefficients in $GdB_6$ is closely related to changes in the AF state. Besides, in $GdB_6$ the absolute values $0.7$-$1.5\times10^{-6}$ $Oe^{-1}$ of the $A_{I,II}$ coefficients (normalized to $T_N$ to be compared also with dodecaborides of the $RB_{12}$ series) are similar to $A_{L,M}(T/T_N) = 1$-$7\times10^{-6}$ $Oe^{-1}$ detected in the $RB_{12}$ family ($R$ – Dy, Ho, Er and Tm) [48, 38]. On the contrary, the values of the negative quadratic coefficient $B_{II}(T/T_N) \sim -(5$-$7)\times10^{-12}$ $Oe^{-2}$ and $B_P(T/T_N) \sim -7\times10^{-12}$ $Oe^{-2}$ detected, correspondingly, in the AF(II) and P-phases of $GdB_6$ are strongly reduced compared to $B_P(RB_{12}) \sim 0.4$-$1.5\times10^{-10}$ $Oe^{-2}$ observed in the paramagnetic state of $RB_{12}$ [38, 48]. Taking into account, that $B_P$ and $B_{II}$ are the characteristics of non-magnetic spin-polarons (heavy fermions) produced by local 4f-5d spin fluctuations [45, 34-37], a significant increase of these parameters along the $RB_{12}$ series may be attributed to enhancement of the quantum mechanical instability of the 4f-shell, which is growing from $DyB_{12}$ towards the mixed valence narrow-gap semiconductor $YbB_{12}$ [36, 49-51]. Moreover, the smallest values $B_P(DyB_{12}) \sim 4\times10^{-11}$ $Oe^{-2}$ have been found in the series of $RB_{12}$ [38] being attributed to development of lattice instability due to proximity of $DyB_{12}$ to the spinodal boundary [39]. The authors [38] argue that the $B_P$ reduction in $DyB_{12}$ results from both very strong lattice dynamics, static displacements and disordering in the non-equilibrium RE dodecaboride. We suppose that similar effects occur in $GdB_6$ where the quantum diffusion regime of charge transport caused by electron and lattice instability was found [23].

It was argued [23] that the quantum diffusion regime arises due to the formation of (i) dynamically coupled $Gd^{3+}$ pairs of about 3.3 Å in size and with energy of quasi-local oscillations ∼7–8 meV, and due to (ii) dynamic charge stripes along the [001] direction of the cubic lattice. Moreover, it has been shown [23] that the anharmonic approximation is appropriate when analyzing the static and dynamic components of the atomic displacement parameters of gadolinium in $GdB_6$. The results of electron spin resonance (ESR) measurements have confirmed the effects related to displacements of $Gd^{3+}$ ions from the centrosymmetric positions in the boron cage of $GdB_6$ [52-53]. Indeed, the easy axis anisotropy with field $H_A \approx 800$ Oe was found in [52], and the mutual displacement $\delta \sim 0.4$ Å of $Gd^{3+}$ ions was estimated. Also strong disordering in location of Gd-ions have been established in the ESR studies of antiferromagnets $Gd_{0.73}La_{0.27}B_6$ and $GdB_6$ [53], concluding in favor of the random mutual shifts of $Gd^{3+}$ spins and formation of the spin-glass state instead of the coherent displacement of Gd ions from the centrally symmetrical positions in the regular cubic lattice.

**III. 5. Jahn-Teller lattice instability and dynamic charge stripes.** Fig. 14a shows the crystal structure model of $GdB_6$, which is typical of most hexaborides. The structure of $GdB_6$ was refined at $T = 30$ K and 85 K in the $Pm\bar{3}m$ group of symmetry and lattice parameter $a_{cub} \approx$ 4.1 Å was deduced. Crystallographic characteristics, details of X-ray diffraction experiment on $GdB_6$ single crystals at 30 K and 85 K and structure refinement results are collected in Table S1 in the Supplementary materials [40]. The cubic lattice distortions $\Delta a \sim 0.008$ Å, $\Delta\varphi \sim 0.15^{\circ}$ are reliably established at 85 K reducing strongly at 30 K (see Table S2 in [40]), however, they are small enough, allowing us to retain the cubic group for this structure model.

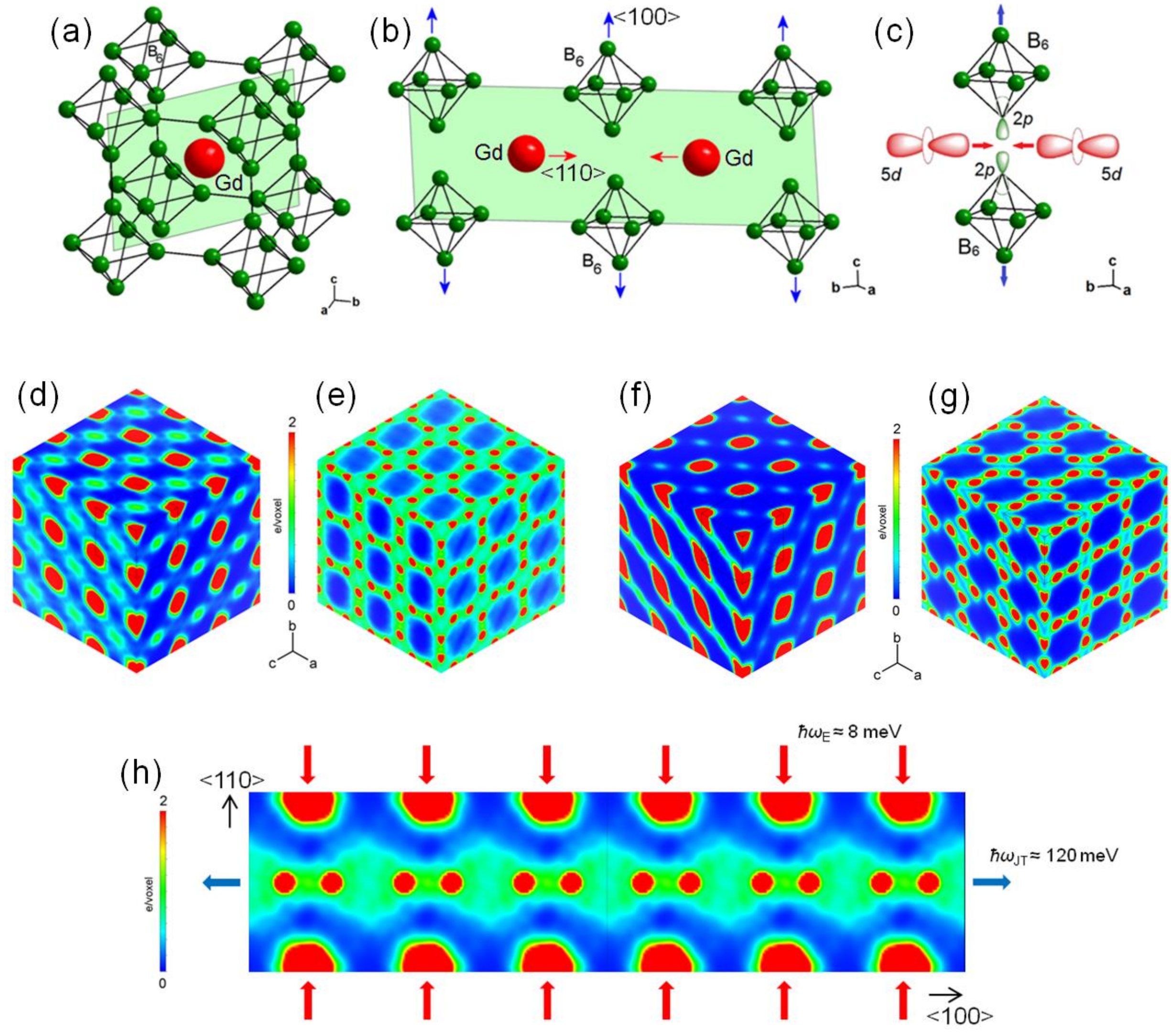

**Fig. 14**. (Color online) (a) Crystal structure of $GdB_6$. (b) Crystal structure in the projection on the (110) plane (this plane is highlighted in green in panels (a) and (b)), blue arrows indicate the vibrations of $B_6$ clusters (the collective JT mode [30]) in one of the directions of the <100> family giving rise to quasi−local vibrations (the Einstein modes [19−23], shown by red arrows) of Gd ions in Gd-Gd pairs in one of the directions of the <110> family, and to periodic changes of 5d−2p hybridization of the band states (see panel (c)). Panels (d)−(e) and (f)–(g) show the electron density distribution in the family of {100} planes at $T$ = 30 K and $T$ = 85 K, correspondingly, obtained by the maximum entropy method (MEM). (h) MEM−constructed map of the ED at $T$ = 30 K with a dynamic charge stripe in one of the {110} planes and the JT boron ($\hbar\omega_{JT} \sim 120$ meV) and Einstein gadolinium ($\hbar\omega_E \sim 8$ meV) modes forming this dynamic stripe (see text). ED peaks were cut off at a height of 2 e/voxel to highlight fine details of the ED distribution in the lattice interstices.

Figs. 14d-14g show the distribution of the electron density (ED) obtained from the same XRD data sets by the maximum entropy method (MEM, see [40] for more detail) at $T$ = 30 K and 85 K in the family of {100} planes, passing through the boron (panels (e) and (g)) and Gd ions ((d) and (f)) in the cubic lattice of $GdB_6$ (see Fig. 14a). The MEM map in one of the {110} planes containing both B and Gd is shown in Fig, 14h. It should be emphasized that ED patterns in three {100} planes, which should be identical in the cubic lattice, are quite different from each other (Figs. S10-S11 in [40] demonstrate also a nonequivalence of ED distribution in the family

of {110} planes). This is due to the combination of two factors: (*i*) the small static JT distortions of the cubic lattice (see Table S2 in [40]) and (*ii*) the cooperative JT dynamic of $B_6$ clusters. The ED distribution patterns in Fig. 14 indicate that at $T$=85 K the 5d-2p dynamic charge stripes are formed in a single direction of the <110> family (Fig. 14f), being localized in the chains constructed from Gd-Gd pairs and bridges of boron atoms. The mechanism of these <110> directed stripes is attributed to the periodical changes of 5d-2p hybridization (Fig. 14c) caused by the collective vibrations of $B_6$ clusters. This electron instability is similar to that observed in the dodecaborides $RB_{12}$ ($R$ = Ho, Er, Tm, Lu), where a 5d-2p fluctuating electron density was found, located strictly along single direction of the <110> family and passing through the *R*- ions and B-B bridges in the spaces between $B_{12}$ clusters [49]. Both the 5d-2p fluctuating electron density developed at 85 K along a single <110> direction and the predominantly 2p-type 3D-grid of stripes detected in $GdB_6$ at $T$=30 K (Fig. 14e) should be attributed to the dynamic JT instability of the boron framework. Note, that JT instability of the boron lattice is a common feature of RE hexaborides, and similar grids of 2p-stripes were observed at $T$ = 30 K also in $LaB_6$ [24-25] and $CeB_6$ [25-28]. A room temperature measurements of wide range dynamic conductivity allowed detecting strong collective modes (overdamped oscillators) in $Gd_xLa_{1-x}B_6$ [29-30], $CeB_6$ [31], $YB_6$ and $YbB_6$ [32]. As shown in [29-32], non-equilibrium (hot) electrons participating in the formation of the collective JT modes dominate in the charge transport of $RB_6$, and the smaller fraction of Drude-type electrons changes in the range 25-45% in the studied hexaborides.

Taking into account, that the Ruderman–Kittel–Kasuya–Yosida (RKKY) indirect exchange is a valid mechanism responsible for the magnetic interaction between the localized magnetic moments of Gd ions, it is natural to expect a strong concurrence between the charge fluctuations in stripes and the RKKY oscillations of spin density of conduction electrons. Therefore, it seems natural to expect (i) the suppression of indirect RKKY exchange between nearest neighbor Gd-ions, and (ii) the *breakdown of cubic magnetic symmetry* in $GdB_6$, with both phenomena (i)-(ii) induced by the charge fluctuations in stripes. Note, that for a number of magnetic RE-dodecaborides $RB_{12}$ (*R*-Dy, Ho, Er, Tm and Yb) similar effects have been established recently [34-38], leading to complicated AF magnetic phase diagrams with numerous magnetic phases and phase transitions. Let us point out finally the relation between the charge stripes along the real space directions <110> and <100>, on the one hand, and two crystal structure modulations with $\boldsymbol{q}_{L1}$= (½, 0, 0) at $T_N$ and $\boldsymbol{q}_{L2}$= (½, ½, 0) superimposed with $\boldsymbol{q}_{L1}$ and developed at $T^*$ ~ 9 K, on the other, while maintaining the same $\boldsymbol{q}_M$ magnetic structure. We propose that instead of the magnetoelastic coupling mechanism with very small (0.004-0.016 Å) displacements of Gd ions [9], one needs to consider the stripe-induced formation of Gd-Gd vibrationally coupled pairs (Fig. 14), which is the second important factor responsible for a strong renormalization of the RKKY-exchange and the cubic symmetry breaking in $GdB_6$.

## IV. Conclusion

A detailed study of a complicated AF state in $GdB_6$ hexaboride with structural (dynamic cooperative Jahn-Teller effect of the boron sub-lattice) and electron instabilities (dynamic charge stripes) was performed using precise angle-dependent measurements of magnetoresistance and magnetization. Comprehensive *H-T, H-φ* and three-dimensional *H-θ-φ* magnetic phase diagrams have been constructed for the first time. The *impeller-type magnetic anisotropy* and the anisotropy of charge carriers scattering is discovered in this antiferromagnetic metal with cubic crystal structure and S-type magnetic ions. The results of precise low-temperature XRD measurements allow concluding in favor of formation of the dynamic charge stripes of two types arranged on: (i) 5d-2p hybridized states of Gd and boron and (ii) a predominantly 2p-orbitals of boron. We propose that the stripe-induced symmetry breaking is accompanied with arrangement of Gd-Gd vibrationally coupled pairs, and these two factors are responsible for the strong renormalization (suppression) of the RKKY indirect exchange interaction.

**Acknowledgements.**

This work was supported by the Russian Science Foundation Project № 26-12-00258, and partly performed using the equipment of the Shared Research Centre of Lebedev Physical Institute of Russian Academy of Sciences and the Center of Excellence of Slovak Academy of Sciences. Authors are grateful to K.M. Krasikov for experimental assistance and A. V. Semeno for helpful discussions. S.G. and K.F. acknowledge the support of the Slovak Research and Development Agency under contract No. APVV−23−0226 and the Slovak Scientific Grant Agency under contract No. VEGA 2/0034/24.

**Supplementary material**
**3D- ($H$-θ-φ) magnetic phase diagram of antiferromagnetic metal $GdB_6$ with electron and lattice instability**

A. N. Azarevich[1], A. V. Bogach[1], T. F. Garipova[1], V. V. Voronov[1], M. Rajnak[2], S. Gabani[2], K. Flachbart[2], N. B. Bolotina[1,3], O. N. Khrykina[1,3], V. M. Gridchina[3,1], A. Yu. Tsvetkov[4], S. Yu. Gavrilkin[4], N. E. Sluchanko[1]

[1]*Prokhorov General Physics Institute, Russian Academy of Sciences, Vavilov str. 38, Moscow 119991, Russia*

[2]*Institute of Experimental Physics of the Slovak Academy of Sciences, Watsonova 47, SK-04001 Košice, Slovakia*
[3]*National Research Center "Kurchatov Institute", Academician Kurchatov sq., 1, Moscow 123182, Russia*
[4]*Lebedev Physical Institute, Russian Academy of Sciences, Leninsky Av. 59, Moscow 119991, Russia*

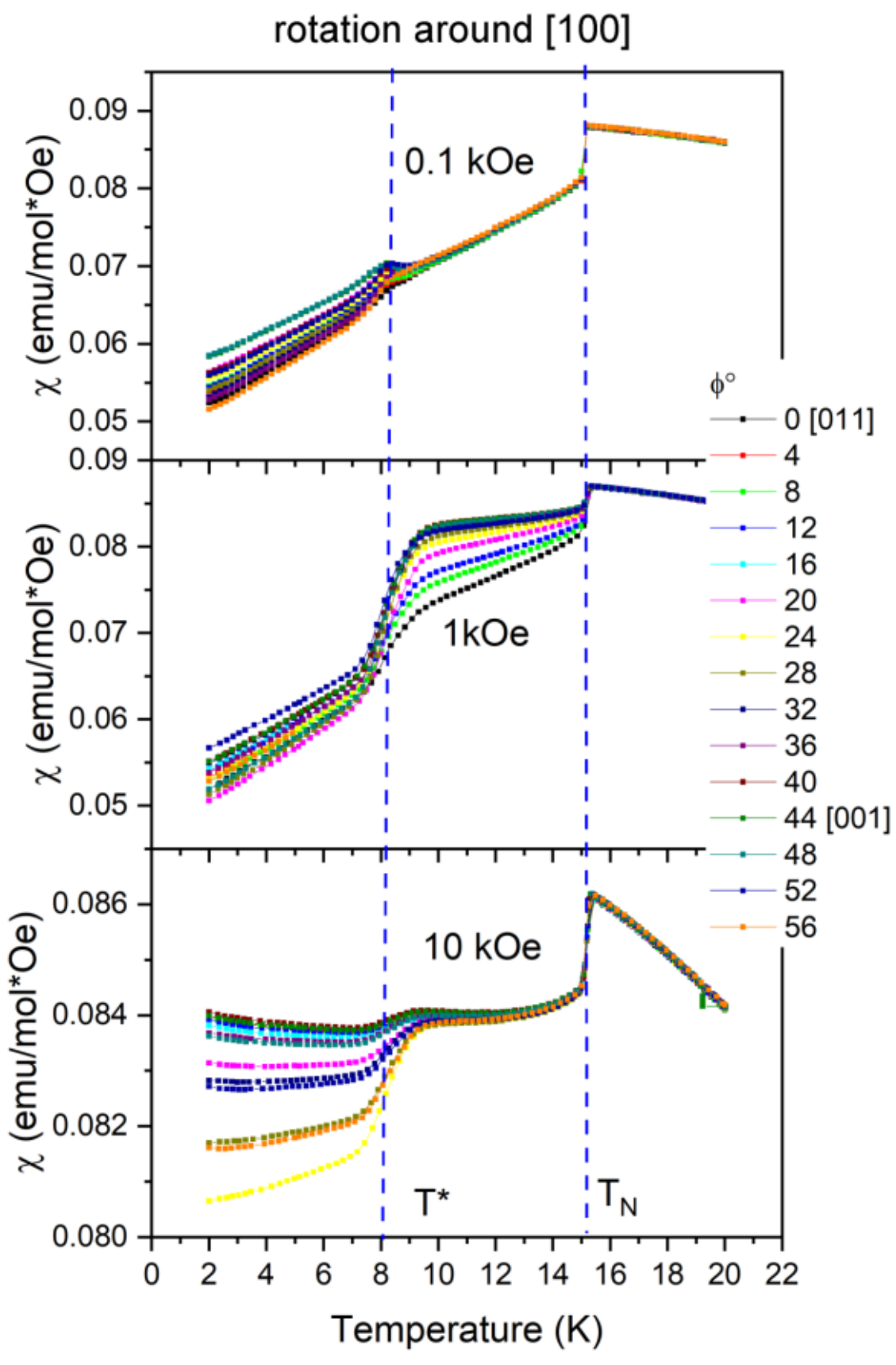

**Fig. S1.** Temperature dependences of magnetic susceptibility at various directions of external magnetic field in the (100) plane with intensity (a) 100 Oe, (b) 1 kOe and (c) 10 kOe. $T_N$ and $T^*$ are the temperatures of magnetic phase transitions, angle $\varphi$ is counted from [011] direction.

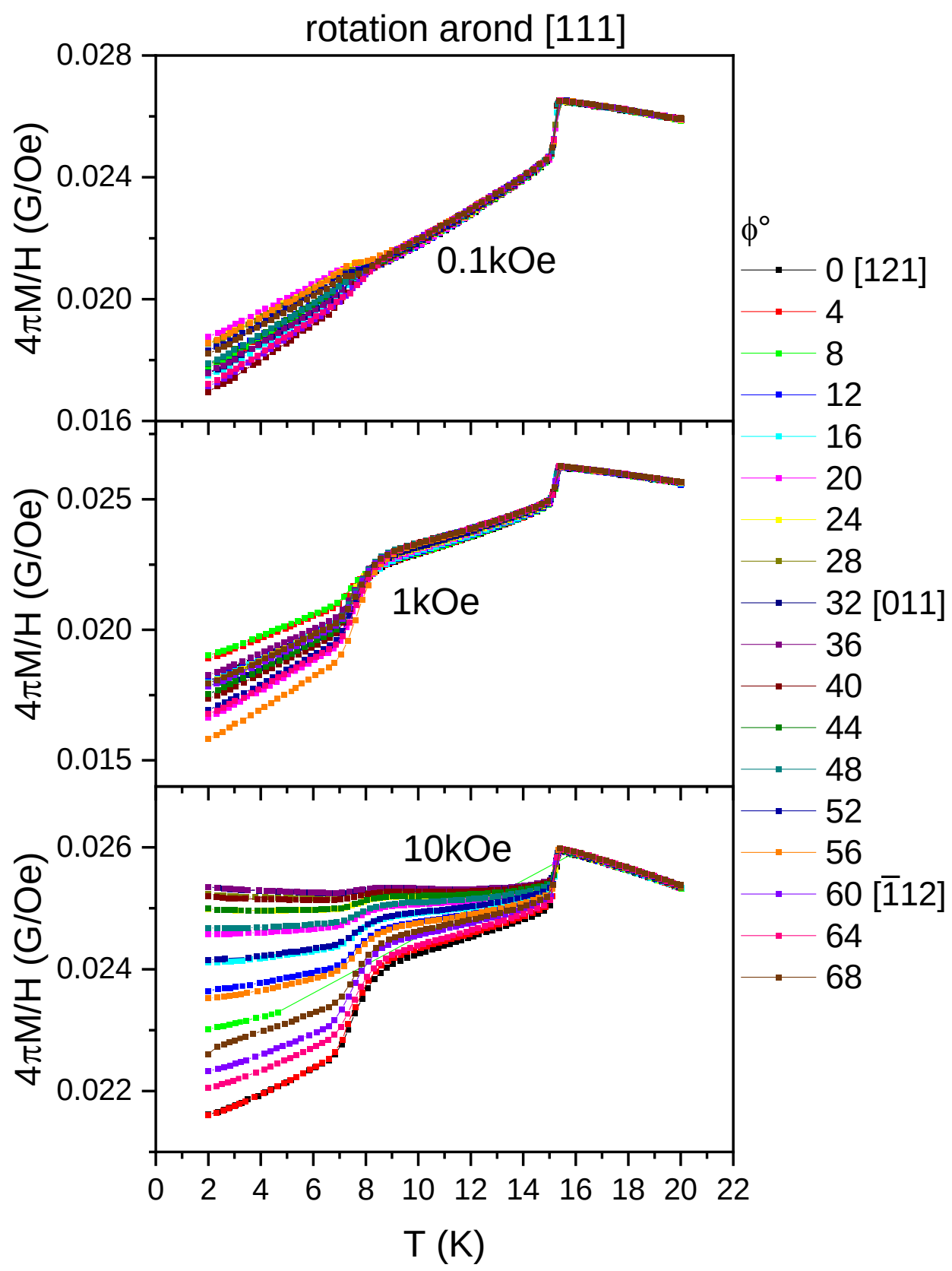

**Fig. S2**. Temperature dependences of magnetic susceptibility at various directions of external magnetic field in the (111) plane with intensity (a) 100 Oe, (b) 1 kOe and (c) 10 kOe. $T_N$ and $T^*$ E are the temperatures of magnetic phase transitions, angle $\varphi$ is counted from [121] direction.

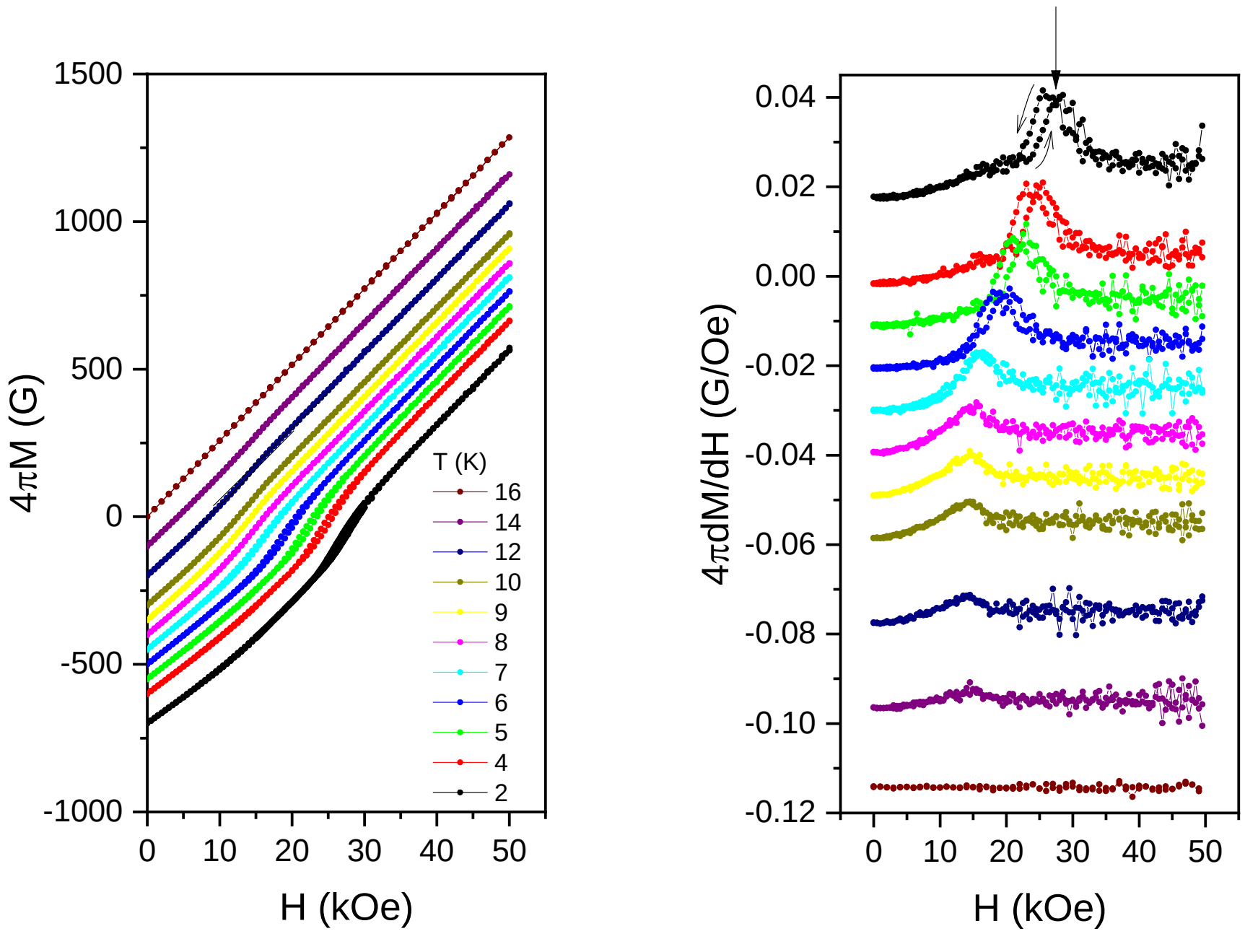

**Fig. S3**. Field dependences of magnetization (a) and the derivative d$M$/d$H$ (b) in the range $H_0 \leq$ 50 kOe at fixed temperatures below 16 K for field direction $\boldsymbol{H}$//[111].

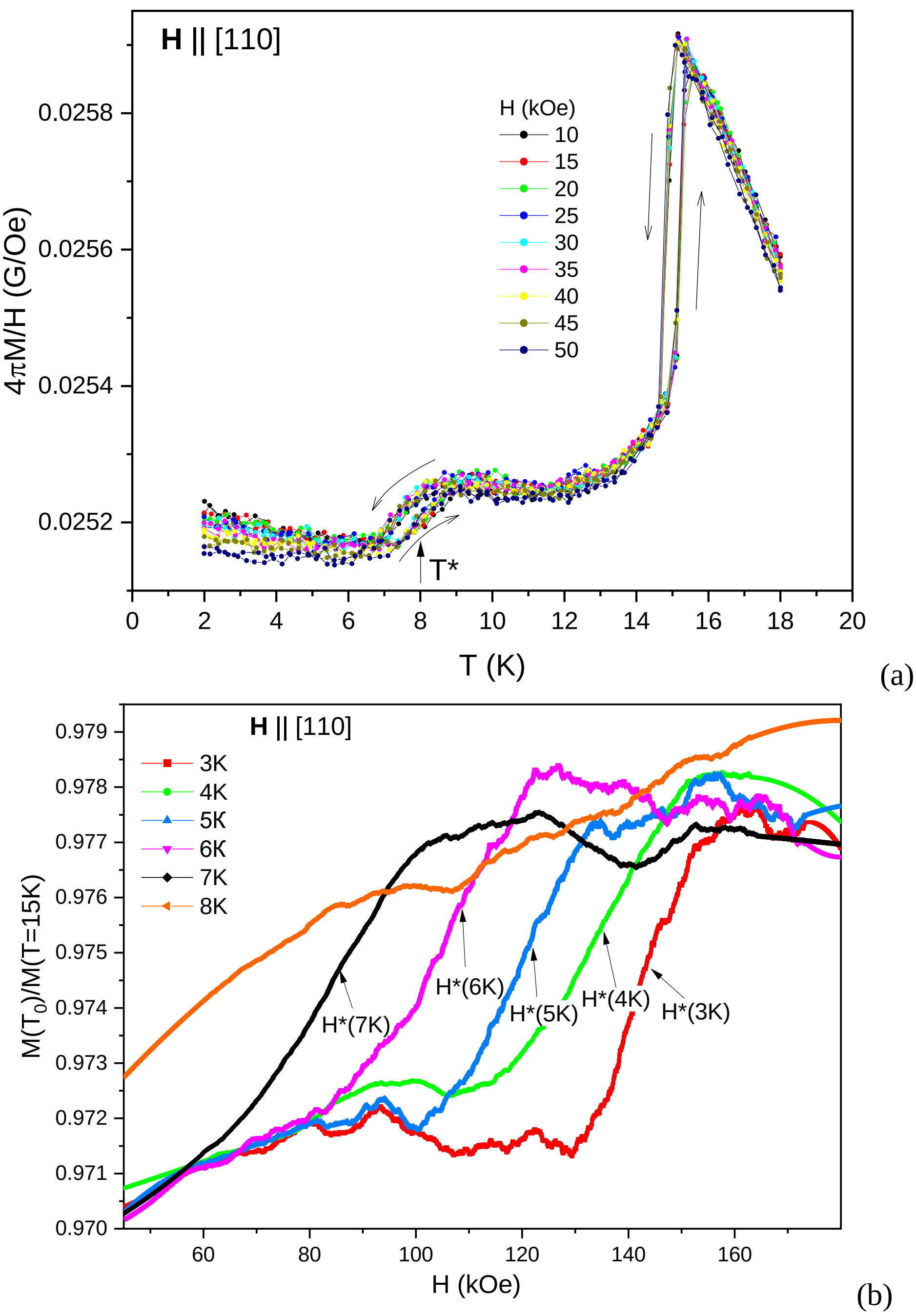

**Fig. S4.** (a) Temperature dependences of magnetic susceptibility $4\pi M/H(T, H_0)$ for $\boldsymbol{H}$//[110] in the range $H_0 \leq 50$ kOe and (b) magnetic field dependences of the normalized magnetization $M(T_0)/M(15\text{ K})$ for $\boldsymbol{H}$//[100] in the range $H_0 \leq 180$ kOe. $T_N$ and $T^*$ are the temperatures of magnetic phase transitions, $H^*$ the critical field of II-I transition.

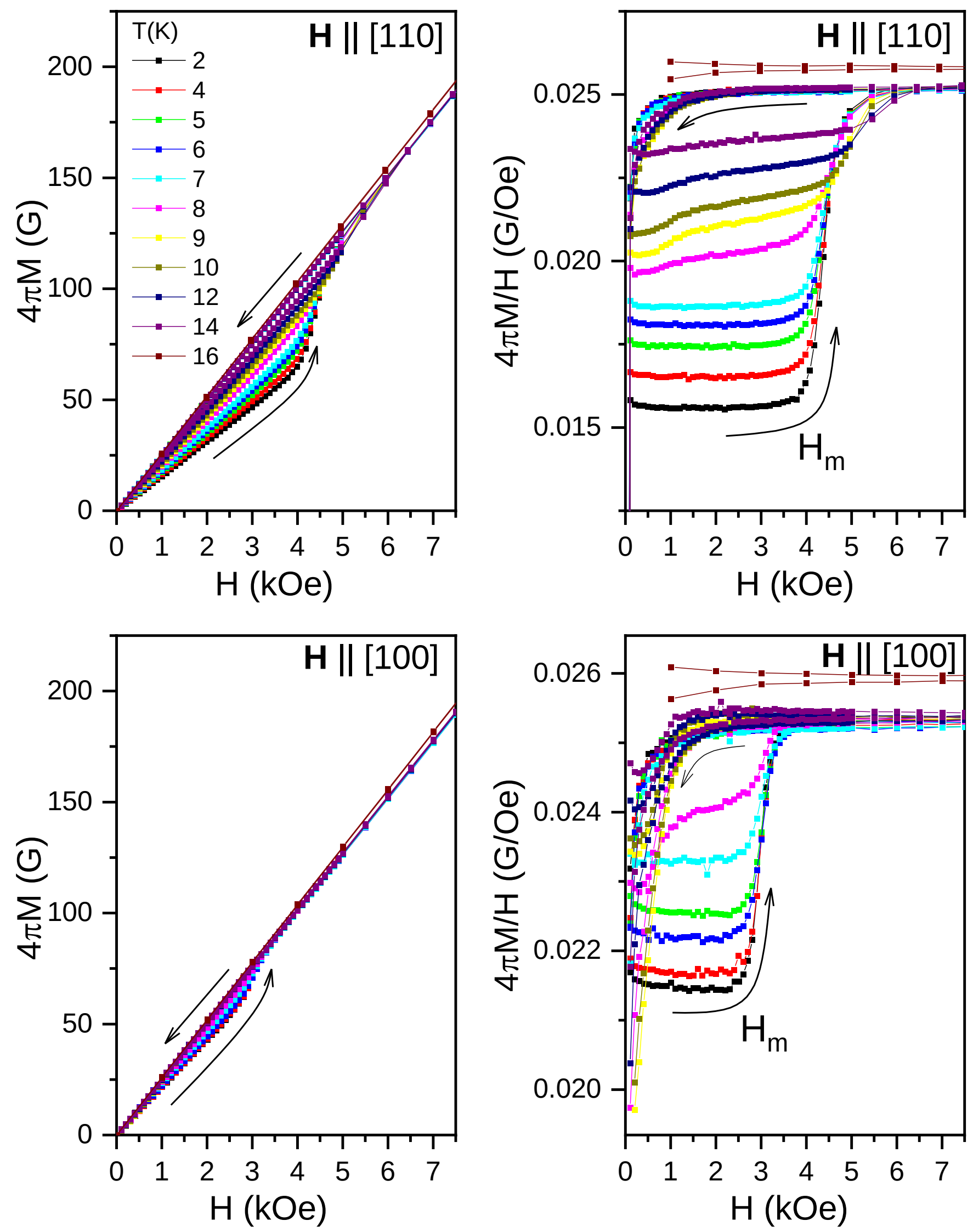

**Fig. S5.** Magnetic field dependences of (a), (c) magnetization $4\pi M(H, T_0)$ and (b), (d) magnetic susceptibility $4\pi M/H(H, T_0)$ for directions ***H***// [110] (a), (b) and ***H***// [100] (c), (d) in the range $H_0 \leq 9$kOe including the hysteresis areas in phases I and II.

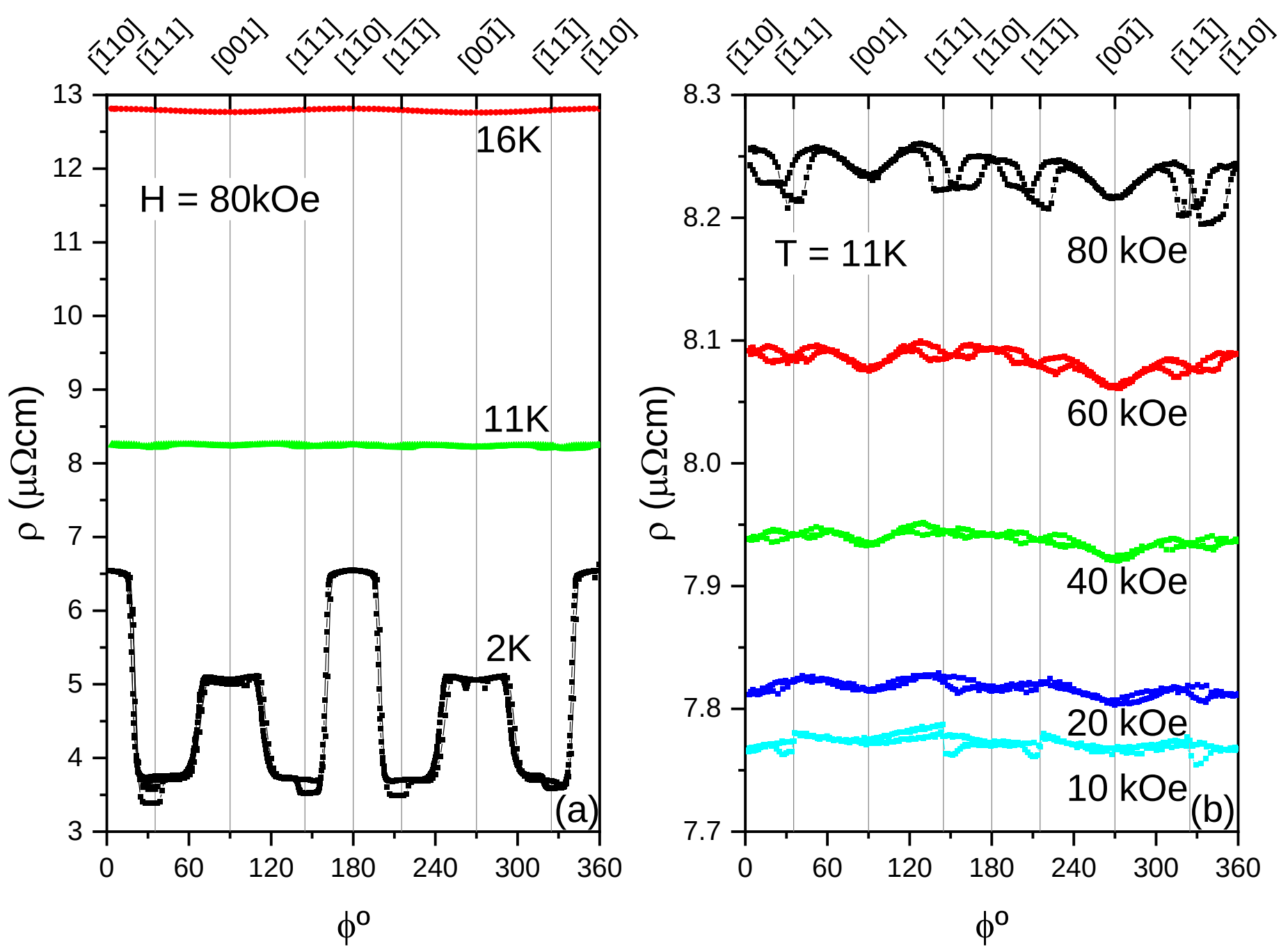

**Fig. S6**. Angular dependences of resistivity for ***H***//(110) (a) at $H$ = 80 kOe and fixed temperatures 2 K (phases I and II), 11 K (phase (I) and 16 K (P-phase), and (b) at $T$ = 11 K and at fixed intensities of external magnetic field in the range 10 - 80 kOe.

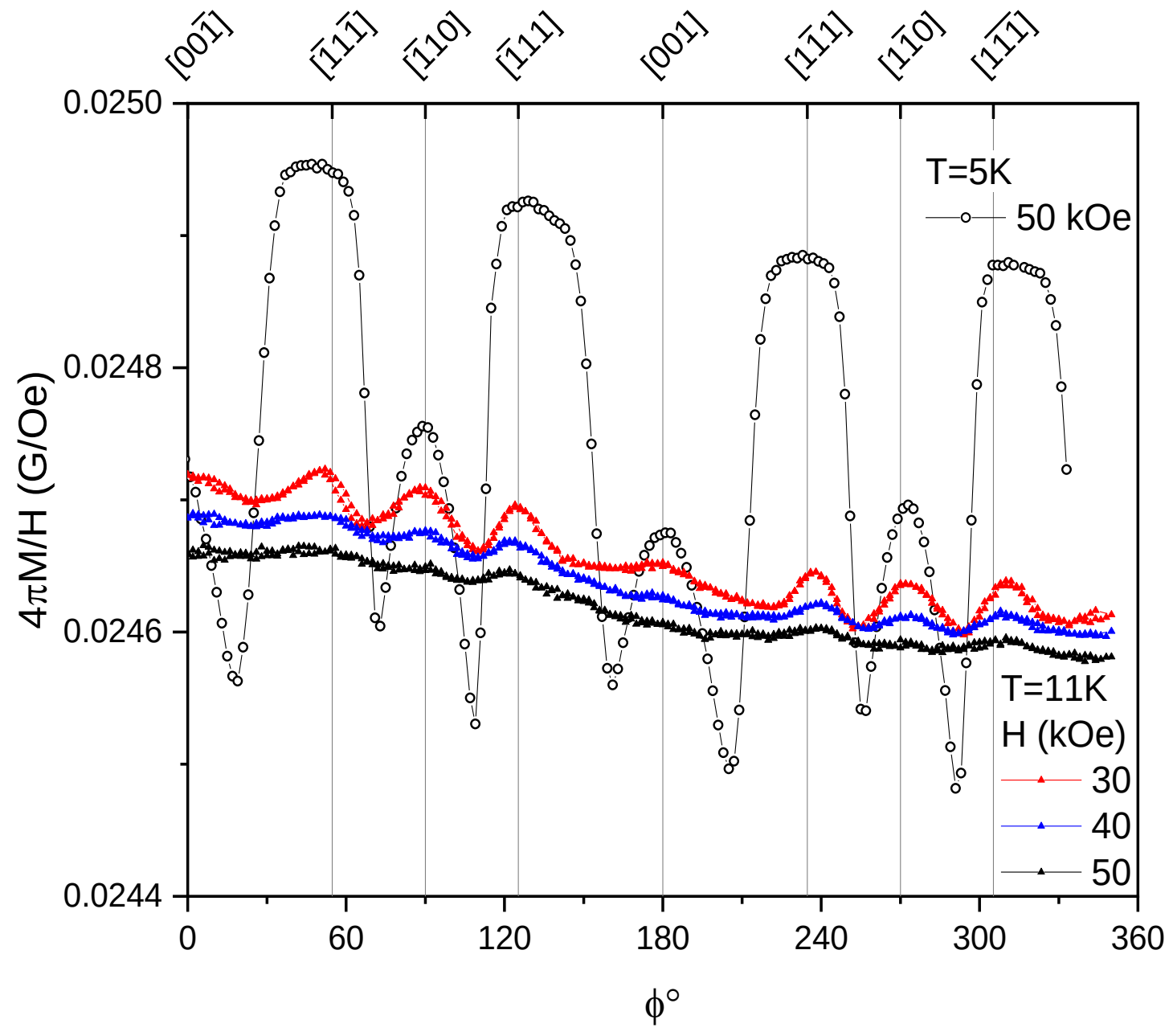

**Fig. S7**. Angular dependences of magnetic susceptibility$4\pi M/H(H, T_0)$ for ***H*** located in the plane (110) recorded at $T_0 = 5$ K and $T_0 = 11$ K when the magnetic field is fixed in the range 30 - 50 kOe. Principal directions in the cubic lattice are denoted by vertical solid lines.

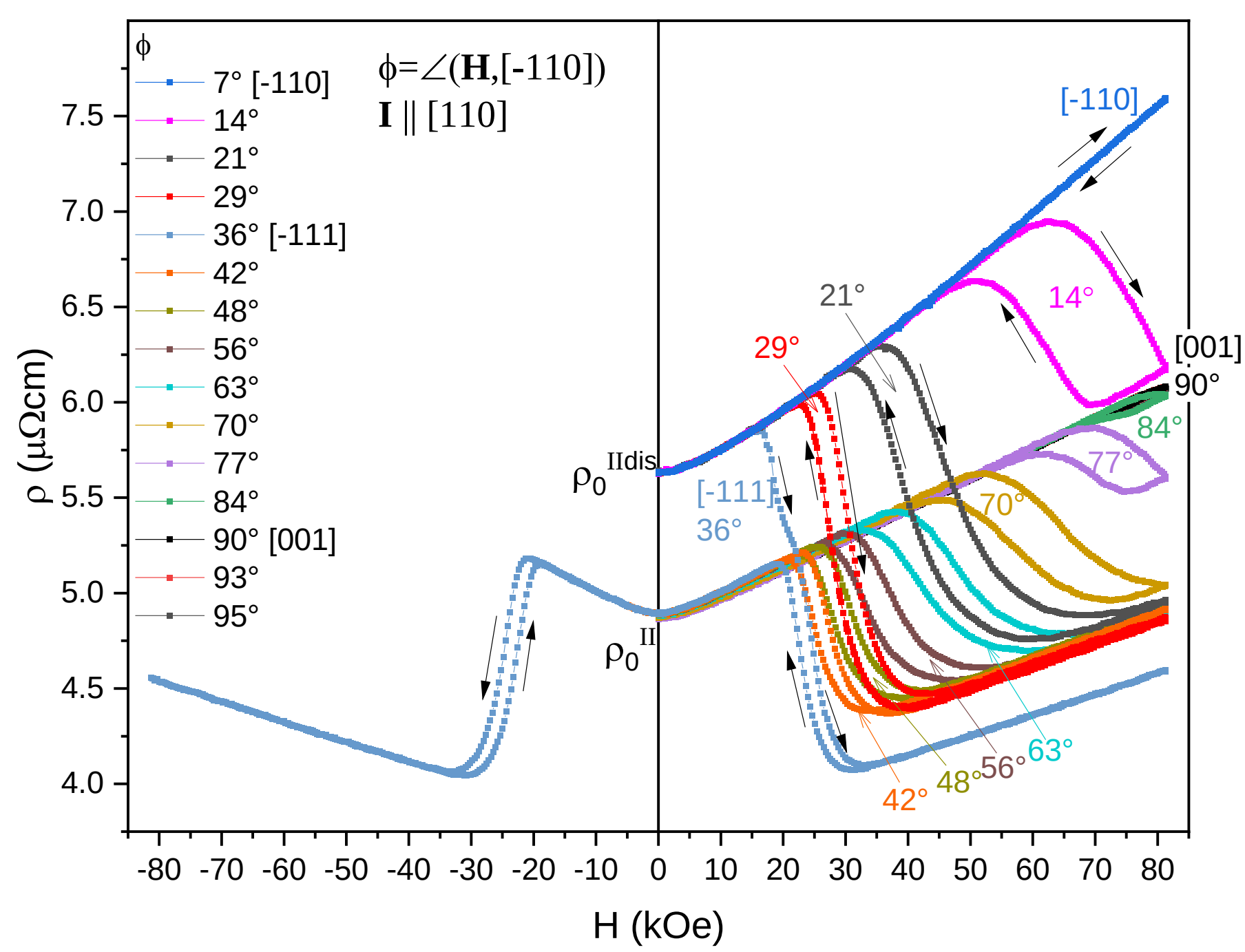

**Fig. S8.** Sweep-up and down magnetic field dependences of resistivity for the various transverse (***H*** ⊥ ***I***) magnetic fields in the range $H \leq 80$ kOe at $T_0 = 5$ K. The step-by-step sample rotation was performed around the direct current axis ***I***//[110]. Up and down directions of field changes are indicated by arrows next to the curves, also showing the hysteresis loop. $\rho_0^{II}$ and $\rho_0^{IIdis}$ demonstrate the disordering-induced difference between zero field resistivity values in phases II and $II_{dis}$.

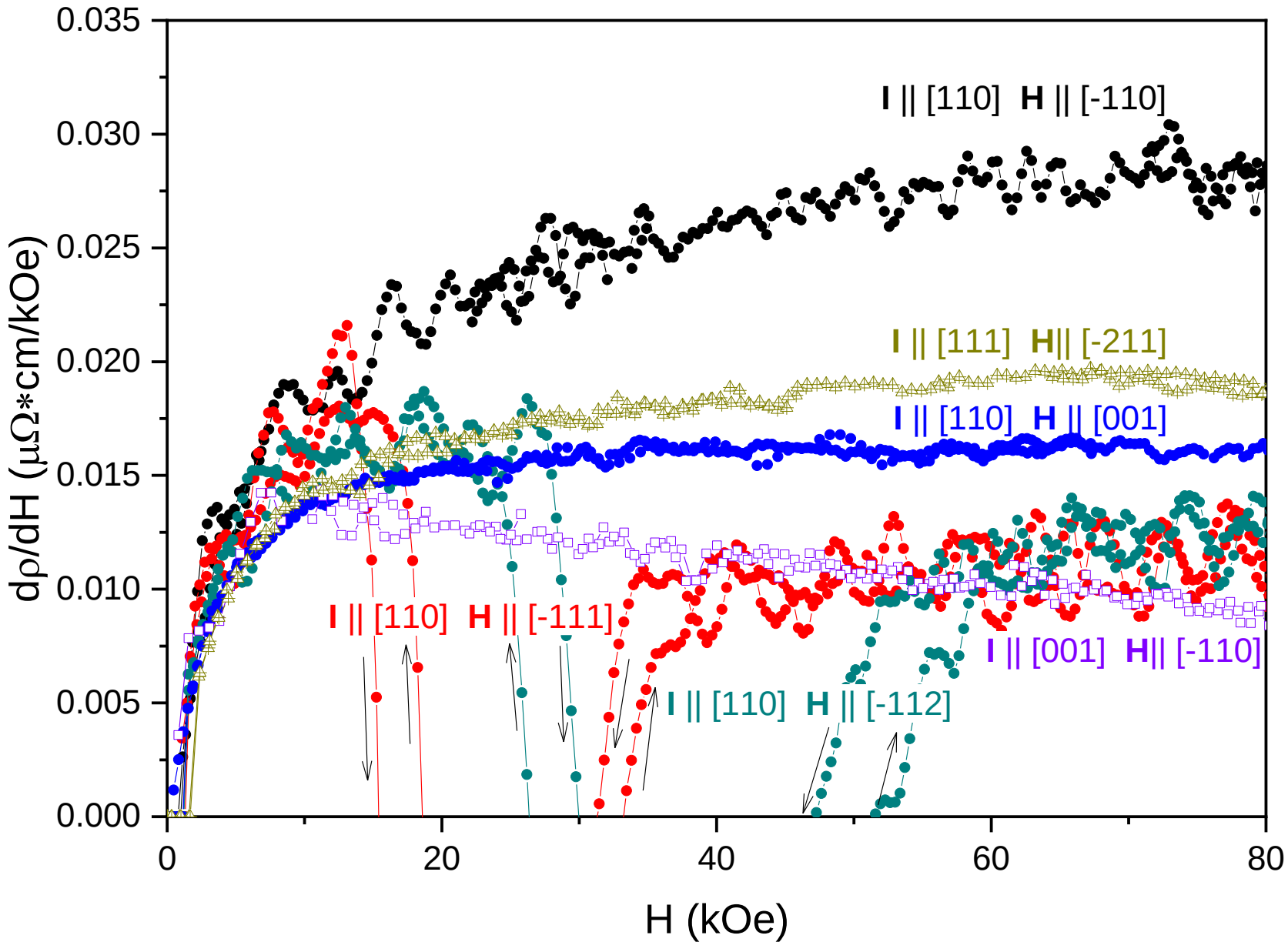

**Fig. S9**. Magnetic field dependences of the resistivity derivatives for various field-current configurations at $T_0 = 5$K. Up and down directions of field sweeps are indicated by arrows next to the curves, also showing the hysteresis loop.

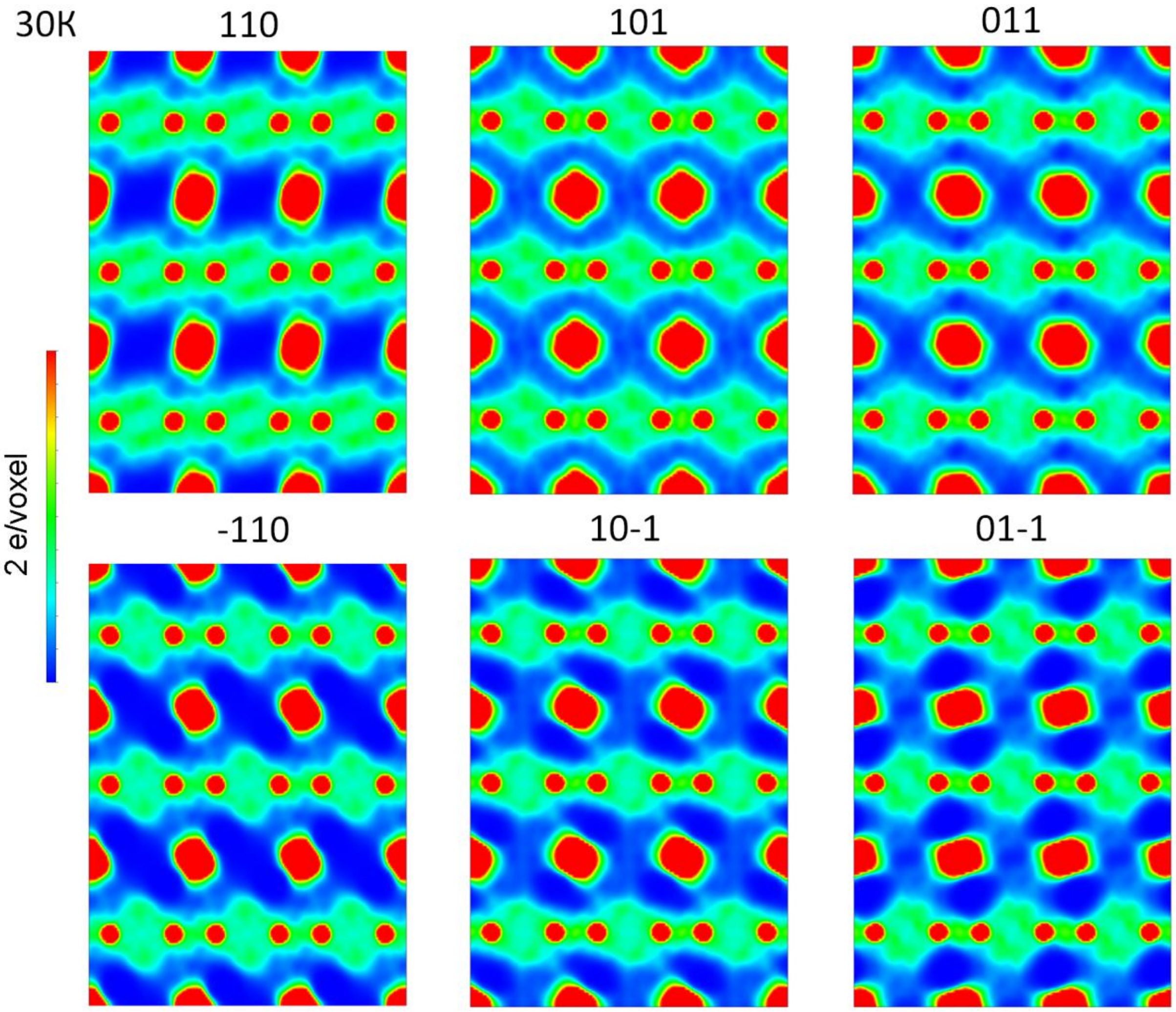

**Fig. S10.** Color maps of the electron density (ED) distribution in the family of {110} planes of $GdB_6$ at $T$ = 30 K obtained using the maximum entropy method (MEM). The calculations were carried out without taking into account the cubic symmetry of the structural model. The ED peaks are truncated at a height of 2 e/voxel to show the fine details of ED distribution in the interstices of the crystal lattice. The plane also contains Gd atoms (large red circles) and B atoms (small red circles).

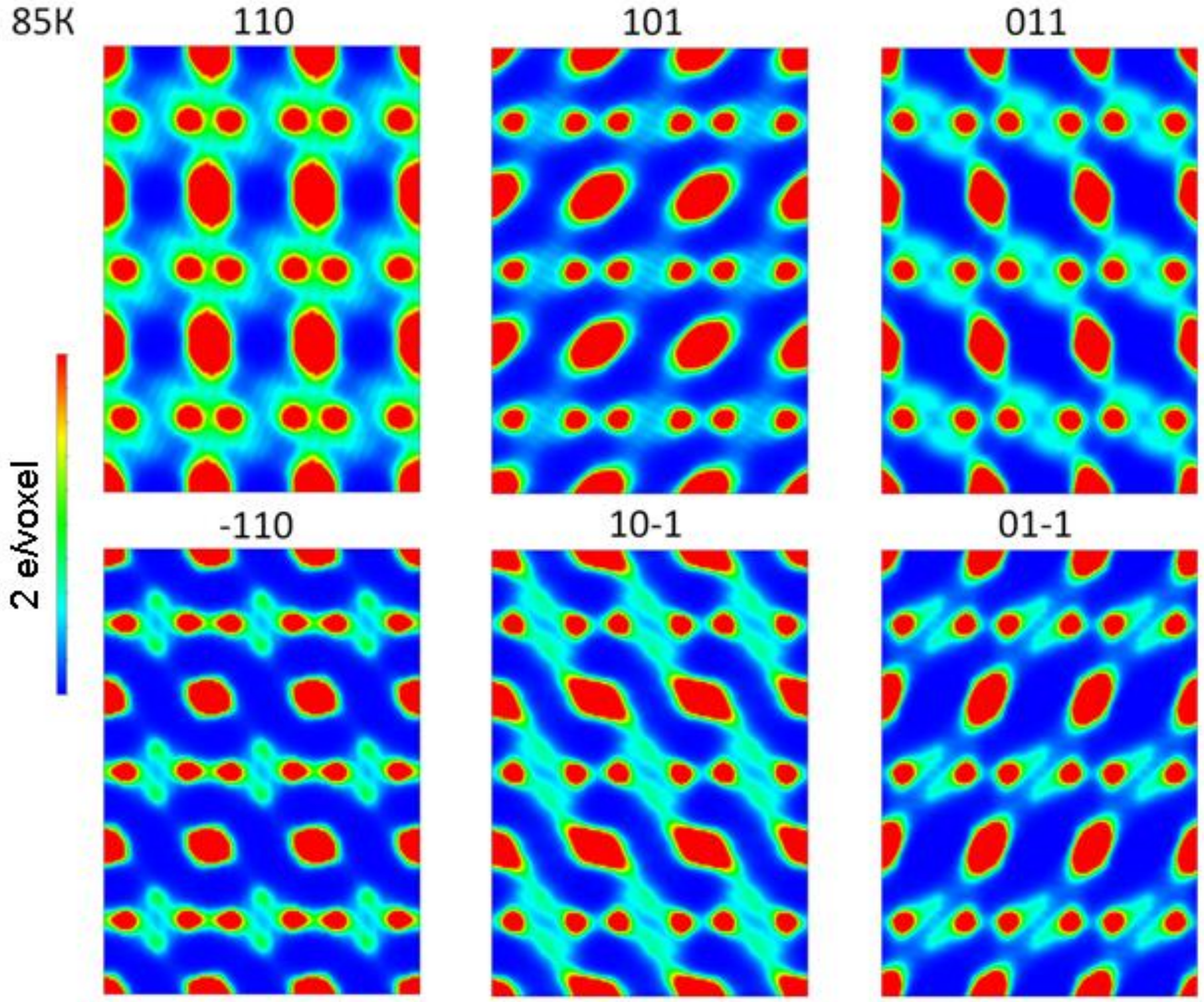

**Fig. S11.** Color maps of the ED distribution in the family of {110} planes of $GdB_6$ at $T$ = 85 K obtained using the maximum entropy method (MEM). The calculations were carried out without taking into account the cubic symmetry of the structural model. The ED peaks are truncated at a height of 2 e/voxel to show the fine details of ED distribution in the interstices of the crystal lattice. The plane also contains Gd atoms (large red circles) and B atoms (small red circles).

**Table S1.** Crystallographic characteristics, details of X-ray diffraction experiments on $GdB_6$ single crystals and structure refinement results.

| Chemical formula | $GdB_6$ | |
|---|---|---|
| $T$, K | 30 | 85 |
| Space group, $Z$ | $Pm$–3$m$, 1 | |
| Lattice parameter $a$, Å | 4.1021 | 4.1000 |
| Radiation type; λ, Å | Ag$K_\alpha$; 0.56087 | Mo$K_\alpha$; 0.71073 |
| Maximum, linear size of the sample, mm | 0.177 | 0.310 |
| Diffractometer | XtaLAB Synergy-DW HyPix-Arc 150° | Xcalibur EosS2 |
| Scan mode | ω | |
| Absorption correction | irregular | sphere |
| Absorption coefficient μ, мм$^{-1}$; $T_{min}$, $T_{max}$ | 12.674; 0.287, 0.540 | 23.708; 0.0188, 0.0991 |
| $\theta_{max}$, deg | 66.99 | 73.56 |
| Limits of $h$; $k$; $l$ | $-13 \le h \le 13$; $-13 \le k \le 13$; $-12 \le l \le 12$ | $-10 \le h \le 10$; $-10 \le k \le 10$; $-10 \le l \le 10$ |
| Number of reflection: observed $N1$; with $I$>3σ$I$; independent $N2$; $R_{int}$, % | 63928; 46764; 314; 6.6 | 7229; 5835; 183; 7.95 |
| $R1_{obs}$/$wR2_{obs}$, % | 0.68/0.81 | 1.50/2.25 |
| $\Delta\rho_{min}$ /$\Delta\rho_{max}$, e/Å$^3$ | –1.71/0.84 | –1.42/1.58 |

**Table S2.** Lattice parameters of $GdB_6$ single crystal without symmetry restrictions.

| *T*, K | 30 | 85 |
|---|---|---|
| *a*, Å | 4.10239(3) | 4.0986(2) |
| *b*, Å | 4.10159(3) | 4.0970(2) |
| *c*, Å | 4.10224(3) | 4.10451(18) |
| α, deg | 89.9783(6) | 90.013(4) |
| β, deg | 89.9756(6) | 89.943(4) |
| γ, deg | 89.9946(6) | 90.092(4) |

**MEM technique**

The Maximum Entropy Method (MEM) is a powerful technique used in various fields of data analysis like signal processing, image analysis, and statistics to estimate the probability distributions or to reconstruct data when limited information is available. It works by finding the probability distribution that is most consistent with the available data (constraints) while being maximally non-committal to any unknown information. This means, it selects the distribution that maximizes entropy, which is a measure of uncertainty or information content. MEM starts with some known information about the system, often expressed as constraints on the expected values of certain functions. These constraints limit the possible probability distributions that can be considered. The method then searches for the probability distribution that satisfies these constraints while having the highest possible entropy. This ensures that the chosen distribution is the least biased and most conservative among all distributions that fit the known information.

When the MEM technique is applied in crystallography, the unit cell is divided into small volumes (voxels) and a MEM-reconstructed electron density (ED) value of $g_{MEM}$ is assigned to each of them [S1]. The method operates directly with the observed and calculated structure factors $F_{obs}$ and $F_{calc}$, but not with atomic coordinates or atomic displacement parameters (ADPs). MEM does not even demand chemical composition of the crystal being guided by general number of electrons in the unit cell. The MEM calculations were made in the present study by Dysnomia program [S2]. The calculations were carried out without taking into account the cubic symmetry of the structural model. The program VESTA [S3] was used for visualization of the MEM results.